\documentclass[twoside,10pt]{article}

\usepackage{jmlr2e}
\usepackage{amsmath,amsfonts,amssymb,amsbsy}
\usepackage{url}
\usepackage{graphicx}
\usepackage{array}
\usepackage{hhline}
\usepackage{dcolumn}
\usepackage{footnote}
\usepackage{xcolor} %
\usepackage{placeins} %
\usepackage{tikz,pgfplots}
\usepackage[titletoc]{appendix} %
\usepackage{natbib} %
\usepackage{booktabs} %
\usepackage{soul} %

\pdfoutput=1
\usepackage{hyperref}
\hypersetup{
     colorlinks=true,
     linkcolor=black,
     citecolor=black,
     filecolor=black,
     urlcolor=black,
}

\def\app#1#2{%
   \mathrel{%
     \setbox0=\hbox{$#1\sim$}%
     \setbox2=\hbox{%
       \rlap{\hbox{$#1\propto$}}%
       \lower1.1\ht0\box0%
     }%
     \raise0.25\ht2\box2%
   }%
}

\usepackage{ifthen}
\usetikzlibrary{matrix}
\usetikzlibrary{calc}
\newlength{\figurewidth}
\newlength{\figureheight}

\def\figpdfdir{./} %
\def\figtikzdir{./} %

\newcommand{\minput}[2][]{
\ifthenelse{\equal{#1}{pdf}}
	{ \includegraphics{\figpdfdir #2} }
	{ \tikzset{external/remake next} \tikzsetnextfilename{#2} \input{\figtikzdir #2} }
}

\usetikzlibrary{external}
\tikzset{external/system call={lualatex
	\tikzexternalcheckshellescape -halt-on-error -interaction=batchmode
	-jobname "\image" "\texsource"}}
\newcommand{\vc}[1] { #1 } %
\newcommand{\vs}[1] { #1 }
\newcommand{\tp}{\mathsf{T}}
\newcommand{\ti}[1] { \tilde{#1} } 
\newcommand{\mc}[1] { \mathcal{#1} } 
\newcommand{\tx}[1] { \text{#1} } 
\newcommand{\given} { \,|\, }
\newcommand{\data} {D}

\newcommand{\pr}[1]{ \Pr {\left[#1\right]} }
\newcommand{\mean}[2][] { \mathrm{E}_{#1} {\left[#2\right]} }
\newcommand{\var}[1] { \mathrm{Var} {\left[#1\right]} }

\newcommand{\KL}[2] { \mathrm{KL} {\left(#1 \, \| \, #2\right)} }

\newcommand{\Normal}[1] { \mathrm{N} {\left(#1\right)}  } 

\newcommand{\halfStudent}[2][] { t_{#1}^+ {\left(#2\right)} }

\newcommand{\InvGamma}[1] { \mathrm{Inv\text{-}Gamma} {\left(#1\right)} }

\newcommand{\halfCauchy}[1] { \mathrm{C}^+ {\left(#1\right)} }
\newcommand{\Beta}[1] {\mathrm{Beta}{\left(#1\right)} }
\newcommand{\Ber}[1] {\mathrm{Ber}{\left(#1\right)} }

\ShortHeadings{Bayesian model selection methods}{Piironen, Vehtari}
\firstpageno{1}

\makeatletter
\def\@starteditor{\noindent \small {}}
\makeatother

\begin{document}

\title{ Comparison of Bayesian predictive methods for\\ model selection }
\author{\name Juho Piironen \email juho.piironen@aalto.fi\\
		\name Aki Vehtari \email aki.vehtari@aalto.fi\\
		\addr Aalto University%
}

\editor{}

\maketitle
\thispagestyle{plain}

\begin{abstract}

The goal of this paper is to compare several widely used Bayesian model selection methods in practical model selection problems, highlight their differences and give recommendations about the preferred approaches.
We focus on the variable subset selection for regression and classification and perform several numerical experiments using both simulated and real world data.
The results show that the optimization of a utility estimate such as the cross-validation (CV) score is liable to finding overfitted models due to relatively high variance in the utility estimates when the data is scarce.
This can also lead to substantial selection induced bias and optimism in the performance evaluation for the selected model.
From a predictive viewpoint, best results are obtained by accounting for model uncertainty by forming the full encompassing model, such as the Bayesian model averaging solution over the candidate models.
If the encompassing model is too complex, it can be robustly simplified by the projection method, in which the information of the full model is projected onto the submodels. 
This approach is substantially less prone to overfitting than selection based on CV-score.
Overall, the projection method appears to outperform also the maximum a posteriori model and the selection of the most probable variables.
The study also demonstrates that the model selection can greatly benefit from using cross-validation outside the searching process both for guiding the model size selection and assessing the predictive performance of the finally selected model.

\end{abstract}
\vspace{0.5\baselineskip}


\section{Introduction}

Model selection is one of the fundamental problems in statistical modeling.
The often adopted view for model selection is not to identify the true underlying model but rather to find a model which is useful.
Typically the usefulness of a model is measured by its ability to make predictions about future or yet unseen observations.
Even though the prediction would not be the most important part concerning the modelling problem at hand, the predictive ability is still a natural measure for figuring out whether the model makes sense or not.
If the model gives poor predictions, there is not much point in trying to interpret the model parameters.
We refer to the model selection based on assessing the predictive ability of the candidate models as {\it predictive model selection}.

Numerous methods for Bayesian predictive model selection and assessment have been proposed and the various approaches and their theoretical properties have been extensively reviewed by \cite{vehtari2012}.
This paper is a follow-up to their work.
The review of \cite{vehtari2012} being qualitative, our contribution is to compare many of the different methods quantitatively in practical model selection problems, discuss the differences, and give recommendations about the preferred approaches.
We believe this study will give useful insights because the comparisons to the existing techniques are often inadequate in the original articles presenting new methods.
Our experiments will focus on variable subset selection on linear models for regression and classification, but the discussion is general and the concepts apply also to other model selection problems.

A fairly popular method in Bayesian literature is to select the maximum a posteriori (MAP) model which, in the case of a uniform prior on the model space, reduces to maximizing the marginal likelihood and the model selection can be performed using Bayes factors \citep[e.g.][]{kass1995, han2001}.
In the input variable selection context, also the marginal probabilities of the variables have been used \citep[e.g.][]{brown1998,barbieri2004,narisetty2014}.
In fact, often in the Bayesian variable selection literature, the selection is assumed to be performed using one of these approaches, and the studies focus on choosing priors for the model space and parameters of the individual models \cite[see the review by][]{ohara2009}.
These studies stem from the fact that sampling the high dimensional model space is highly nontrivial, and because it is well known that both the Bayes factors and the marginal probabilities are sensitive to the prior choices \citep[e.g.][]{kass1995,ohara2009}.

However, we want to make a distinction between prior and model selection.
More specifically, we want to address the question, given our prior beliefs, how should we choose the model?
In our view, the prior choice should reflect our genuine beliefs about the unknown quantities, such as the number of relevant input variables.
For this reason our goal is not to compare and review the vast literature on the priors and computation strategies, which is already done by \cite{ohara2009}.
Still, we feel this literature is very essential, giving tools for formulating different prior beliefs and performing the necessary posterior computations.

In other than variable selection context, a common approach is to choose the model based on its estimated predictive ability, preferably by using Bayesian leave-one-out cross-validation (LOO-CV) \citep{geisser1979} or the widely applicable information criterion (WAIC) \citep{watanabe2009book}, both of which are known to give a nearly unbiased estimate of the predictive ability of a given model \citep{watanabe2010}.
Also several other predictive criteria with different loss functions have been proposed, for instance the deviance information criterion (DIC) by \cite{spiegelhalter2002} and the various squared error measures by \cite{laud1995}, \cite{gelfand1998}, and \cite{marriott2001} none of which, however, are unbiased estimates of the true generalization utility of the model.

Yet an alternative approach is to construct a full encompassing reference model which is believed to best represent the uncertainties regarding the modelling task and choose a simpler submodel that gives nearly similar predictions as the reference model.
This approach was pioneered by \cite{lindley1968} who considered input variable selection for linear regression and used the model with all the variables as the reference model.
The idea was extended by \cite{brown1999,brown2002}.
A related method is due to \cite{sanmartini1984} who used the Bayesian model averaging (BMA) solution as the reference model and Kullback--Leibler divergence for measuring the difference between the predictions of the reference model and the candidate model instead of the squared error loss used by Lindley.
Another related method is the reference model projection by \cite{goutis1998} and \cite{dupuis2003} in which the information contained in the posterior of the reference model is projected onto the candidate models.
The variations of this method include heuristic Lasso-type searching \citep{nott2010} and approximative projection with different cost functions \citep{tran2012, hahn2015}.

Although LOO-CV and WAIC can be used to obtain a nearly unbiased estimate of the predictive ability of a given model, both of these estimates contain a stochastic error term whose variance can be substantial when the dataset is not very large.
This variance in the estimate may lead to overfitting in the selection process causing nonoptimal model selection and inducing bias in the performance estimate for the selected model \citep[e.g.][]{ambroise2002,reunanen2003,cawley2010}.
The overfitting in the selection may be negligible if only a few models are being compared but, as we will demonstrate, may become a problem for a larger number of candidate models, such as in variable selection.

Our results show that the optimization of CV or WAIC may yield highly varying results and lead to selecting a model with predictive performance far from optimal.
From the predictive point of view, best results are generally obtained by accounting for the model uncertainty and forming the full BMA solution over the candidate models, and one should not expect to do better by selection.
Our results agree with what is known about the good performance of the BMA \citep{hoeting1999,raftery2003}.
If the full model is too complex or the cost for observing all the variables is too high, the model can be simplified most robustly by the projection method which is considerably less vulnerable to the overfitting in the selection.
The advantage of the projection approach comes from taking into account the model uncertainty in forming the full encompassing model and reduced variance in the performance evaluation of the candidate models.
Overall, the projection framework outperforms also the selection of the most probable inputs or the MAP model, while both of these typically perform better than optimization based on CV or WAIC.

Despite its advantages, the projection approach has suffered from a difficulty in deciding how many variables should be selected in order to get predictions close to the reference model \citep{peltola2014, robert2014}.
Our study shows that the model selection can highly benefit from using cross-validation outside the variable searching process both for guiding the model size selection and assessing the predictive performance of the finally selected model.
Using cross-validation for choosing only the model size rather than the variable combination introduces substantially less overfitting due to greatly reduced number of model comparisons (see Sec.~\ref{sec:modelsize_selection} for discussion).

\begin{table*}%
\centering
\caption{Categorization of the different model selection methods discussed in this paper. }
\label{tab:selection_methods}
\abovetopsep=2pt
\begin{tabular}{l  l  l  }
\toprule
 & View & Methods \\
\midrule 
Section~\ref{sec:utility_estimation} &
Generalization utility estimation ($\mc M$-open view)  &
Cross-validation, WAIC, DIC  \\ \noalign{\smallskip}
Section~\ref{sec:other_criteria} &
Self/posterior predictive criteria (Mixed view)  & 
$L^2$-, $L^2_\tx{cv}$- and $L^2_k$-criteria  \\ \noalign{\smallskip}
Section~\ref{sec:reference_methods} &
Reference model approach  ($\mc M$-completed view)  &
\parbox[t]{4.5cm}{Reference predictive method,\\ Projection predictive method}  \\ \noalign{\smallskip}
Section~\ref{sec:model_space} &
Model space approach ($\mc M$-closed view)   &
\parbox[t]{4.5cm}{Maximum a posteriori model, Median probability model}  \\
\bottomrule
\end{tabular}
\end{table*}

The paper is organized as follows.
In Section~\ref{sec:methods} we shortly go through the model selection methods discussed in this paper.
This part of the paper somewhat overlaps with the previous studies \citep[especially with][]{vehtari2012}, but is included for maintaining easier readibility as a standalone paper.
Section~\ref{sec:selection_bias} discusses and illustrates the overfitting in model selection and the consequent selection induced bias in the performance evaluation of the chosen model.
Section~\ref{sec:results} is devoted to the numerical experiments and forms the core of the paper.
The discussion in Section~\ref{sec:conclusions} concludes the paper.

\section[Methods]{Approaches for Bayesian model selection}
\label{sec:methods}

This section discusses shortly the proposed methods for Bayesian model selection relevant for this paper.
We do not attempt anything close to a comprehensive review but summarize the methods under comparison.
See the review by \cite{vehtari2012} for a more complete discussion.\sloppy

The section is organized as follows.
Section~\ref{sec:expected_utility} discusses how the predictive ability of a model is defined in terms of an expected utility and Sections~\ref{sec:utility_estimation}--\ref{sec:model_space} shortly review the methods.
Following \cite{bernardo1994book} and \cite{vehtari2012} we have categorized the methods into  $\mc M$-closed, $\mc M$-completed, and $\mc M$-open views, see Table~\ref{tab:selection_methods}.
$\mc M$-closed means assuming that one of the candidate models is the true data generating model. %
Under this assumption, one can set prior probabilities for each model and form the Bayesian model averaging solution (see Section~\ref{sec:model_space}).
The $\mc M$-completed view abandons the idea of a true model, but still forms a reference model which is believed to be the best available description of the future observations.
In the $\mc M$-open view one does not assume one of the models being true and also rejects the idea of constructing the reference model. 

Throughout Section~\ref{sec:methods}, the notation assumes a model $M$ which predicts an output $y$ given an input variable $x$.
The same notation is used both for scalars and vectors.
We denote the future observations by $\ti y$ and the model parameters by $\vs \theta$.
To make the notation simpler, we denote the training data as $\data = \{(x_i,y_i)\}_{i=1}^n$.

\subsection{Predictive ability as an expected utility}
\label{sec:expected_utility}

The predictive performance of a model is typically defined in terms of a utility function that describes the quality of the predictions.
An often used utility function measuring the quality of the predictive distribution of the candidate model $M$ is the logarithmic score \citep{good1952}
\begin{align}
	u(M,\ti y) = \log p(\ti y \given \data, M) \,.
\end{align}
Here we have left out the future input variables $\vc{\ti x}$ to simplify the notation.
The logarithmic score is a widely accepted utility function due to its information theoretic grounds and will be used in this paper.
However, we point out that in principle any other utility function could be considered, and the choice of a suitable utility function might also be application specific.

Since the future observations $\ti y$ are unknown, the utility function $u(M,\ti y)$ cannot be evaluated beforehand.
Therefore one usually works with the expected utilities instead 
\begin{align}
	\bar u(M) = \mean{\log p(\ti y \given \data, M)} = \int p_\tx{t}(\ti y) \log p(\ti y \given \data, M) \mathrm d \ti y,
\label{eq:util_gen}
\end{align}
where $p_\tx{t}(\ti y)$ denotes the true data generating distribution.
This expression will be referred to as the generalization utility or more loosely as the predictive performance of model $M$.
Maximizing~\eqref{eq:util_gen} is equivalent to minimizing the Kullback--Leibler (KL) divergence from the true data generating distribution $p_\tx{t}(\ti y)$ to the predictive distribution of the candidate model $M$.

\subsection{Generalization utility estimation}
\label{sec:utility_estimation}

\subsubsection{Cross-validation}
The generalization utility~\eqref{eq:util_gen} can be estimated by using the already obtained sample $\data$ as proxy for the true data generating distribution $p_\tx{t}(\ti y)$.
However, estimating the expected utility using the same data $\data$ that were used to train the model leads to an optimistic estimate of the generalization performance.
A better estimate is obtained by dividing the data into $K$ subsets $I_1,\dots,I_K$ and using each of these sets in turn for validation while training the model using the remaining $K-1$ sets.
This gives the Bayesian $K$-fold cross-validation (CV) utility \citep{geisser1979}
\begin{align}
	K\tx{-fold-CV} 
	= \frac{1}{n} \sum_{i=1}^n \log p(y_i \given \vc x_i, \data_{\setminus I_{s(i)}}, M),
\label{eq:kfcv}
\end{align}
where $I_{s(i)}$ denotes the validation set that contains the $i$th point and $D_{\setminus I_{s(i)}}$ the training data from which this subset has been removed.
Conditioning the predictions on fewer than $n$ data points introduces bias in the utility estimate.
This bias can be corrected \citep{burman1989} but small $K$ increases the variance in the estimate. 
One would prefer to set $K=n$ computing the leave-one-out utility (LOO-CV) but without any computational shortcuts this is often computationally infeasible as the model would need to be fitted $n$ times.
An often used compromise is $K = 10$.
Analytical approximations for LOO are discussed by \cite{vehtari2014a} and computations using posterior samples  by \cite{vehtari2014b}.

\subsubsection{Information criteria}
Information criteria offer a computationally appealing way of estimating the generalization performance of the model.
A fully Bayesian criterion is the widely applicable information criterion (WAIC) by \cite{watanabe2009book, watanabe2010}.
WAIC can be calculated as
\begin{align}
	\tx{WAIC} = \frac{1}{n} \sum_{i=1}^n \log p(y_i \given \vc x_i, \data, M) - \frac{V}{n},
\label{eq:waic}
\end{align}
where the first term is the training utility and $V$ is the functional variance given by
\begin{align}
	V = \sum_{i=1}^n \left\lbrace \mean{(\log p(y_i \given \vc x_i,\vs \theta, M))^2}
	  					-\mean{\log p(y_i \given \vc x_i,\vs \theta, M)}^2
	  				\right\rbrace \,.
\end{align}
Here both of the expectations are taken over the posterior $p(\vs \theta \given \data,M)$.
\cite{watanabe2010} proved that WAIC is asymptotically equal to the Bayesian LOO-CV and to the generalization utility~\eqref{eq:util_gen}, and the error is $o(1/n)$.
Watanabe's proof gives Bayesian LOO and WAIC a solid theoretical justification in the sense that they measure the predictive ability of the model including the uncertainty in the parameters and can be used also for singular models (the set of the ``true parameters'' consists of more than one point).

Another still popular method is the deviance information criterion (DIC) proposed by \cite{spiegelhalter2002}.
DIC estimates the generalization performance of the model with parameters fixed to the posterior mean $\vs{\bar \theta} = \mean{\vs \theta \given \data, M}$.
DIC can be written as
\begin{align}
	\tx{DIC} = \frac{1}{n} \sum_{i=1}^n \log p(y_i \given \vc x_i, \vs{\bar \theta}, M)
			 - \frac{p_\tx{eff}}{n},
\label{eq:dic}
\end{align}
where $p_\tx{eff}$ is the effective number of parameters which can be estimated as
\begin{align}
	p_\tx{eff} = 2 \sum_{i=1}^n \left( \log p(y_i \given \vc x_i, \vs{\bar \theta},M)
				- \mean{ \log p(y_i \given \vc x_i, \vs \theta, M) } \right),
\label{eq:dic_peff1}
\end{align}
where the expectation is taken over the posterior.
From Bayesian perspective, DIC is not theoretically justified since it measures the fit of the model when the parameters are fixed to the posterior expectation and is not therefore an unbiased estimate of the true generalization utility~\eqref{eq:util_gen}.
The use of a point estimate is questionable not only because of Bayesian principles, but also from a practical point of view especially when the model is singular.

\subsection{Mixed self and posterior predictive criteria}
\label{sec:other_criteria}

There exists a few criteria that are not unbiased estimates of the true generalization utility~\eqref{eq:util_gen} but have been proposed for model selection.
These criteria do not fit to the $\mc M$-open view since the candidate models are partially assessed based on their own predictive properties and therefore these criteria resemble $\mc M$-closed/$\mc M$-completed view \citep[for a detailed discussion, see][]{vehtari2012}. 

\cite{ibrahim1994} and \cite{laud1995} proposed a selection criterion for regression derived by considering replicated measurements $\ti{\vc y}$ at the training inputs.
The criterion measures the expected squared error between the new observations and the old ones $\vc y$ over the posterior predictive distribution of the candidate model $M$.
The error measure can be decomposed as
\begin{align}
	L^2 &= \sum_{i=1}^n (y_i - \mean{\ti y \given \vc x_i, \data, M})^2 
			+ \sum_{i=1}^n \var{\ti y \given \vc x_i, \data, M}, 
\label{eq:L2}
\end{align}
that is, as a sum of the squared errors for mean predictions plus sum of the predictive variances.
The preferred model is then the one which minimizes~\eqref{eq:L2}.
$L^2$-criterion is not an unbiased estimate of~\eqref{eq:util_gen} due to different form of the utility function, but \cite{ibrahim1994} showed that in model comparison, the criterion penalizes more complex models asymptotically with penalty halfway between the posterior predictive approach (i.e. fit at the training points) and the cross-validation approach.

\cite{marriott2001} proposed a closely related criterion which is a cross-validated version of~\eqref{eq:L2}
\begin{align}
	L^2_\tx{cv} = \sum_{i=1}^n (y_i - \mean{\ti y \given \vc x_i, \data_{\setminus I_{s(i)}}, M})^2 
			+ \sum_{i=1}^n \var{\ti y \given \vc x_i, \data_{\setminus I_{s(i)}}, M} \,.
\label{eq:L2cv}
\end{align}
This sounds intuitively better than the $L^2$-criterion because it does not use the same data for training and testing.
However, little is known about the properties of the estimate~\eqref{eq:L2cv} as the authors do not provide a theoretical treatment.
Empirically it is found that, like $L^2$, $L^2_\tx{cv}$-criterion has a relatively high variance which may cause significant overfitting in model selection as discussed in Section~\ref{sec:selection_bias} and demonstrated experimentally in Section~\ref{sec:results}.

Yet another related criterion based on a replicated measurement was proposed by \cite{gelfand1998}.
The authors considered an optimal point prediction which is designed to be close to both the observed and future data and the relative importance between the two is adjusted by a free parameter $k$.
Omitting the derivation, the loss function becomes
\begin{align}
	L^2_k = \frac{k}{k+1} \sum_{i=1}^n (y_i - \mean{\ti y \given \vc x_i, \data, M})^2
			 + \sum_{i=1}^n \var{\ti y \given \vc x_i, \data, M} \,.
\label{eq:L2k}
\end{align}
When $k \rightarrow \infty$, this is the same as the $L^2$-criterion~\eqref{eq:L2}.
On the other hand, when $k=0$, the criterion reduces to the sum of the predictive variances.
In this case there is no inherent safeguard against poor fit as the model with the narrowest predictive distribution is chosen.
In their experiment, the authors reported that the results were not very sensitive to the choice of $k$.

\subsection{Reference model approach}
\label{sec:reference_methods}

Section~\ref{sec:utility_estimation} reviewed methods for utility estimation that are based on sample reuse without any assumptions on the true model ($\mc M$-open view).
An alternative way is to construct a full encompassing reference model $M_*$, which is believed to best describe our knowledge about the future observations, and perform the utility estimation almost as if it was the true data generating distribution ($\mc M$-completed view).
We refer to this as the reference model approach.
There are basically two somewhat different but related approaches that fit into this category, namely the reference and projection predictive methods, which will be discussed separately.

\subsubsection{Reference predictive method}
\label{sec:ref_pred}
Assuming we have constructed a reference model $M_*$ which we believe best describes our knowledge about the future observations, the utilities of the candidate models $M$ can be estimated by replacing the true distribution $p_\tx{t}(\ti y)$ in~\eqref{eq:util_gen} by the predictive distribution of the reference model $p(\ti y \given \data, M_*)$.
Averaging this over the training inputs $\{\vc x_i\}_{i=1}^n$ gives the reference utility
\begin{align}
\label{eq:refutil}
	\bar u_\tx{ref} = \frac{1}{n}\sum_{i=1}^n \int p(\ti y \given \vc x_i, \data, M_*)
									 \log p(\ti y \given \vc x_i, \data, M) \mathrm d \ti y \,.
\end{align}
Depending on the reference model, this integral may not be analytically available and numerical integration may be required.
However, if the output $\ti y$ is only one dimensional, simple quadratures are often adequate.

As the reference model is in practice different from the true data generating model, the reference utility is a biased estimate of the true generalization utility~\eqref{eq:util_gen}.
The maximization of the reference utility is equivalent to minimizing the predictive KL-divergence between the reference model $M_*$ and the candicate model $M$ at the training inputs
\begin{align}
\label{eq:ref_discrepancy}
	\delta(M_* \| M) 
	& = \frac{1}{n} \sum_{i=1}^n
	 \KL{p(\ti y \given \vc x_i, \data, M_*)}{p(\ti y \given \vc x_i, \data, M)}.
\end{align}
The model choice can then be based on the strict minimization of the discrepancy measure~\eqref{eq:ref_discrepancy}, or choosing the simplest model that has an acceptable discrepancy.
What is meant by ``acceptable'' may be somewhat arbitrary and depend on the context.
For more discussion, see the concept of relative explanatory power in the next section, Equation~\eqref{eq:explanatory_power}.

The reference predictive approach is inherently a less straightforward approach to model selection than the methods presented in Section~\ref{sec:utility_estimation}, because it requires the construction of the reference model and it is not obvious how it should be done.
\cite{sanmartini1984} proposed using the Bayesian model average~\eqref{eq:bma} as the reference model (see Sec.~\ref{sec:model_space}).
In the variable selection context, the model averaging corresponds to a spike-and-slab type prior \citep{mitchell1988} which is often considered as the ``gold standard'' and has been extensively used for linear models \citep[see, e.g.,][]{george1993, george1997, raftery1997, brown2002, lee2003, ohara2009, narisetty2014} and extended and applied to regression for over one million variables \citep{peltola2012a,peltola2012b}.
However, we emphasize that any other model or prior could be used as long as we believe it reflects our best knowledge of the problem and allows convenient computation.
For instance the Horseshoe prior \citep{carvalho2009,carvalho2010} has been shown to have desirable properties empirically and theoretically assuming a properly chosen shrinkage factor \citep{datta2013,varDerPas2014}.

\subsubsection{Projection predictive method}
\label{sec:projection}

A related but somewhat different alternative to the reference predictive method (previous section) is the projection approach.
The idea is to project the information in the posterior of the reference model $M_*$ onto the candidate models $M$ so that the predictive distribution of the candidate model remains as close to the reference model as possible.
Thus the key aspect is that the candidate model parameters are determined by the fit of the reference model, not by the data.
Therefore also the prior needs to be specified only for the reference model.
The idea behind the projection is quite generic and \cite{vehtari2012} discuss the general framework in more detail.

A practical means for doing the projection was proposed by \cite{goutis1998} and further discussed by \cite{dupuis2003}.
Given the parameter of the reference model $\vs \theta^*$, the projected parameter $\vs \theta^\perp$ in the parameter space of model $M$ is defined via 
\begin{align}
	\vs \theta^\perp 
	&= \arg \min_{\vs \theta} \frac{1}{n}\sum_{i=1}^n
						\KL{p(\ti y \given \vc x_i, \vs \theta^*, M_*)} 
							{p(\ti y \given \vc x_i, \vs \theta, M)} \,.
\label{eq:projection}
\end{align}
The discrepancy between the reference model $M_*$ and the candidate model $M$ is then defined to be the expectation of the divergence over the posterior of the reference model
\begin{align}
	\delta(M_* \| M)
	= \frac{1}{n}\sum_{i=1}^n
	\mean%
	{ \KL{p(\ti y \given \vc x_i, \vs \theta^*, M_*)}
		{p(\ti y \given \vc x_i, \vs \theta^\perp, M)} } \,.
\label{eq:projection_discrepancy}
\end{align}
The posterior expectation in~\eqref{eq:projection_discrepancy} is in general not available analytically.
\cite{dupuis2003} proposed calculating the discrepancy by drawing samples $\{\vs \theta_s^*\}_{s=1}^S$ from the posterior of the reference model, calculating the projected parameters $\{\vs \theta_s^\perp\}_{s=1}^S$ individually according to~\eqref{eq:projection}, and then approximating~\eqref{eq:projection_discrepancy} as
\begin{align}
	\delta(M_* \| M) \approx
	\frac{1}{n S}\sum_{i=1}^n \sum_{s=1}^S
	\KL{p(\ti y \given \vc x_i, \vs \theta_s^*, M_*)}
		{p(\ti y \given \vc x_i, \vs \theta_s^\perp, M)} \,.
\label{eq:projection_discrepancy_approx}
\end{align}
Also the optimization problem in~\eqref{eq:projection} cannot typically be solved analytically, and a numerical optimization routine may be needed.
However, for the simplest models like the Gaussian linear model, the analytical solution is available, see Appendix~\ref{app:proj_lgm}.
Moreover, even when the analytical solution does not exist, solving the optimization problem~\eqref{eq:projection} in the case of generalized linear models is equivalent to finding the maximum likelihood parameters for the candidate model $M$ with the observations replaced by the fit of the reference model \citep{goutis1998}.

The projected samples $\{\vs \theta_s^\perp\}_{s=1}^S$ are used for posterior inference as usual.
For example, the predictions of the candidate model $M$ can be computed as
\begin{align}
	p(\ti y \given \ti x, D, M) = \frac{1}{S} \sum_{s=1}^S p(\ti y \given \ti x, \vs \theta_s^\perp, M),
\label{eq:projection_pred}
\end{align}
which is the same as the usual Monte Carlo approximation to the predictive distribution, we simply use the projected samples as the posterior approximation. 

For deciding which model model to choose, \cite{dupuis2003} introduced a measure called relative explanatory power
\begin{align}
	\phi(M) = 1- \frac{\delta(M_* \| M)} {\delta(M_* \| M_0)} \,,
\label{eq:explanatory_power}
\end{align}
where $M_0$ denotes the empty model, that is, the model that has the largest discrepancy to the reference model.
In terms of variable selection, $M_0$ is the variable free model.
By definition, the relative explanatory power obtains values between 0 and 1, and \cite{dupuis2003} proposed choosing the simplest model with enough explanatory power, for example 90\%, but did  not discuss the effect of this threshold for the predictive performance of the selected models.
We note that, in general, the relative explanatory power is an unreliable indicator of the predictive performance of the submodel.
This is because the reference model is typically different from the true data generating model $M_\tx{t}$, and therefore both $M_*$ and $M$ may have the same discrepancy to $M_\tx{t}$ (that is, the same predictive ability) although the discrepancy between $M_*$ and $M$ would be nonzero.

\cite{peltola2014} proposed an alternative way of deciding the model size based on cross-validation outside the searching process.
In other words, in a $K$-fold setting the searching is repeated $K$ times each time leaving $1/K$ of the data for testing, and the performance of the found models are tested on this left-out data.
Note that also the reference model is trained $K$ times and each time its performance is evaluated on the left-out data.
Thus, one can compare the utility of both the found models and the reference model on the independent data and estimate, for instance, how many variables are needed to get statistically indistinguishable predictions compared to the reference model.
More precisely, if $u_m$ denotes the estimated expected utility of choosing $m$ variables and $u_*$ denotes the estimated utility for the reference model, the models can be compared by estimating the probability
\begin{align}
	\pr{u_*  - u_m \le  \Delta u},
\label{eq:pr_u_diff}
\end{align}
that is, the probability that the utility difference compared to the reference model is less than $\Delta u \ge 0$.
\cite{peltola2014} suggested estimating the probabilities above by using Bayesian bootstrap \citep{rubin1981} and reported results for all model sizes for $\Delta u = 0$.

The obvious drawback in this approach are the increased computations (as the selection and reference model fitting is repeated $K$ times), but in Section~\ref{sec:results_real} we demonstrate that this approach may be very useful when choosing the final model size.

\subsection{Model space approach}
\label{sec:model_space}

Bayesian formalism has a natural way of describing the uncertainty with respect to the used model specification given an exhaustive list of candidate models $\{M_\ell\}_{\ell=1}^L$.
The distribution over the model space is given by
\begin{align}
	p(M \given \data) \propto  p(\data \given M) \, p(M) \,.
\label{eq:model_prob}
\end{align}
The predictions are then obtained from the Bayesian model averaging (BMA) solution
\begin{align}
\label{eq:bma}
	p(\ti{y} \given \data) = \sum_{\ell=1}^L p(\ti y \given \data, M_\ell)\, p(M_\ell \given \data) \,.
\end{align}
Strictly speaking, forming the model average means adopting the $\mc M$-closed view, that is, assuming one of the candidate models is the true data generating model.
In practice, however, averaging over the discrete model space does not differ in any sense from integrating over the continuous parameters which is the standard procedure in Bayesian modeling.
Moreover, BMA has been shown to have a good predictive performance both theoretically and empirically \citep{raftery2003} and especially in variable selection context the integration over the different variable combinations is widely accepted.
See the review by \cite{hoeting1999} for a thorough discussion of Bayesian model averaging.

From a model selection point of view, one may choose the model maximizing~\eqref{eq:model_prob} ending up with the maximum a posteriori (MAP) model. 
Assuming the true data generating model belongs to the set of the candidate models, MAP model can be shown to be the optimal choice under the zero-one utility function (utility being one if the true model is found, and zero otherwise).
If the models are given equal prior probabilities, $p(M)\propto 1$, finding the MAP model reduces to maximizing the marginal likelihood, also referred to as the type-II maximum likelihood.

\cite{barbieri2004} proposed a related variable selection method for the Gaussian linear model and named it the Median probability model (which we abbreviate simply as the Median model).
The Median model is defined as the model containing all the variables with marginal posterior probability greater than $\frac{1}{2}$.
Let binary vector $\vs \gamma=(\gamma^1,\dots,\gamma^p)$ denote which of the variables are included in the model ($\gamma^j=1$ meaning that variable $j$ is included).
The marginal posterior inclusion probability of variable $j$ is then
\begin{align}
	\pi_j = \sum_{M \,:\, \gamma^j = 1} p(M \given \data),
\end{align}
that is, the sum of the posterior probabilities of the models which include variable $j$.
The Median model $\vs \gamma_{\text{med}}$ is then defined componentwise as
\begin{align}
	\gamma^j_{\text{med}} = 
	\begin{cases}
	1, \quad \text{if}\thickspace \pi_j \ge \frac{1}{2}, \\
	0, \quad \text{otherwise} \,.
	\end{cases}
\label{eq:median_model}
\end{align}
The authors showed that when the predictors $\vc x = (x^1,\dots,x^p)$ are orthogonal, that is when $\vc Q = \mean {\vc x \vc x^\tp}$ is diagonal, the Median model is the optimal choice.
By optimal the authors mean the model whose predictions for future $\ti y$ are closest to the Bayesian model averaging prediction~\eqref{eq:bma} in the squared error sense.
The authors admit that the assumption of the orthogonal predictors is a strong condition that does not often apply.
The Median model also assumes that the optimality is defined in terms of the mean predictions, meaning that the uncertainty in the predictive distributions is ignored.
Moreover, the Median model is derived assuming Gaussian noise and thus the theory does not apply, for instance, to classification problems.

\begin{figure*}[t]
\centering
	\footnotesize
	\setlength{\figureheight}{0.23\textwidth}
	\setlength{\figurewidth}{0.44\textwidth}
	\pgfplotsset{
	compat=newest,
	title style={yshift=-0.8em, text height=2em, font=\normalsize}
	}
	\minput[pdf]{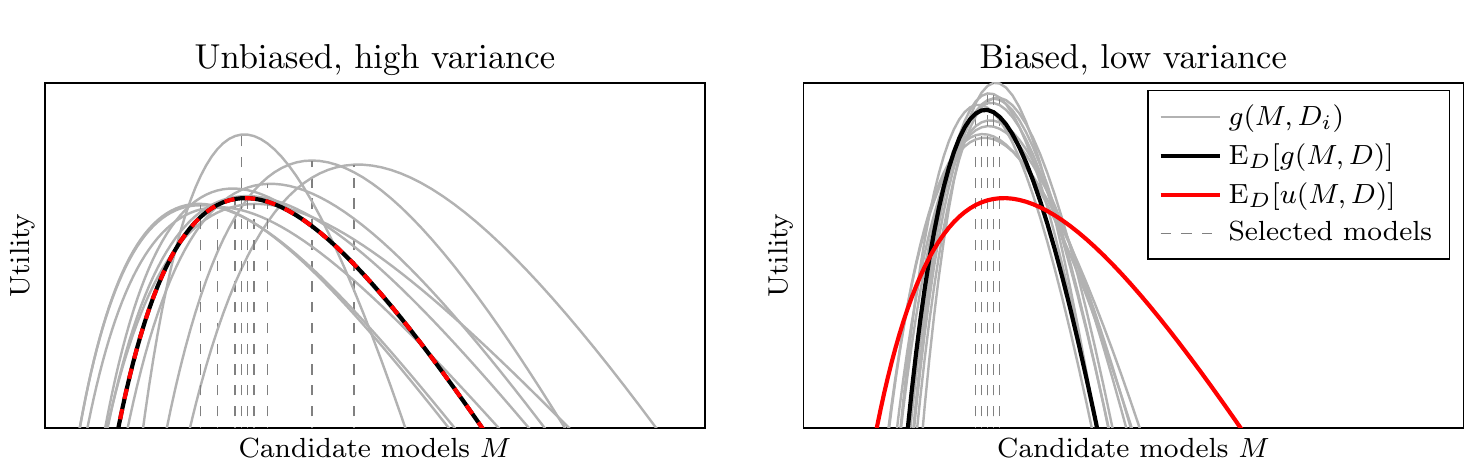}
	\caption{Schematic illustration of an unbiased (left) and a biased (right) utility estimation method. Grey lines denote the utility estimates for different datasets $\data_i$, black is the average, and red the true expected utility. In this case, the biased method is likely to choose better models (dashed lines) due to better tradeoff in bias and variance.}
  \label{fig:selectionbias}
\end{figure*}

\section[Overfitting in selection]{Overfitting and selection induced bias}
\label{sec:selection_bias}

As discussed in Section~\ref{sec:expected_utility}, the performance of a model is usually defined in terms of the expected utility~\eqref{eq:util_gen}.
Many of the proposed selection criteria rewieved in Sections~\ref{sec:utility_estimation}--\ref{sec:model_space} can be thought of as estimates of this quantity even if not designed directly for this purpose.

Consider a hypothetical utility estimation method.
For a fixed training dataset  $\data$, its utility estimate $g_\ell = g(M_\ell,\data)$ for model $M_\ell$ can be decomposed as
\begin{align}
	g_\ell = u_\ell + e_\ell,
\label{eq:util_decomp}
\end{align}
where $u_\ell = u(M_\ell, \data)$ represents the true generalization utility of the model, and $e_\ell = e(M_\ell, \data)$ is the error in the utility estimate.
Note that also $u_\ell$ depends on the observed dataset $\data$, because favourable datasets lead to better generalization performance.
If the utility estimate is correct on average over the different datasets $\mean[\data]{g_\ell } = \mean[\data]{u_\ell}$, or equivalently $\mean[\data]{e_\ell} = 0$, we say the estimate $g$ is unbiased, otherwise it is biased.
The unbiasedness of the utility estimate is often considered as beneficial for model selection.
However, the unbiasedness is intrinsically unimportant for model selection, and a successful model selection does not necessarily require unbiased utility estimates.
To see this, note that the only requirement for a perfect model selection criterion is that the higher utility estimate implies higher generalization performance, that is $g_\ell > g_k$ implies $u_\ell > u_k$ for all models $M_\ell$ and $M_k$.
This condition can be satisfied even if $\mean[\data]{g_\ell} \ne \mean[\data]{u_\ell}$.

To get an idea how the bias and variance properties of a utility estimate affect the model selection, see Figure~\ref{fig:selectionbias}.
The left plot shows an imaginary prototype of an unbiased but high variance utility estimation method.
The grey lines represent the estimated utilities for each model $M$ with different data realizations.
On average (black) these curves coincide with the true expected utility over all datasets (red).
However, due to the high variance, the maximization of the utility estimate may lead to choosing a model with nonoptimal expected true utility (the maxima become scattered relatively far away from the true optimum).
We refer to this phenomenon of choosing a nonoptimal model due to the variance in the utility estimates as {\it overfitting} in model selection.
In other words, the selection procedure fits to the noise in the utility estimates and therefore it is expected that the chosen model has a nonoptimal true utility.
The left plot also demonstrates that, even though the utility estimates are unbiased for each model before the selection, the utility estimate for the selected model is no longer unbiased and is typically optimistic (the maxima of the grey lines tend to lie over the average curve).
We refer to this as the {\it selection induced bias}.

The right plot shows a biased utility estimation method that either under or overestimates the ability of most of the models.
However, due to smaller variance, the probability of choosing a model with better true performance is significantly increased (the maxima of the estimates focus closer to the true optimum).
This example demonstrates that even though the unbiasedness is beneficial for the performance evaluation of a particular model, it is not necessarily important for model selection.
For the selection, it is  more important to be able to rank the competing models in an approximately correct order with a low variability.

The overfitting in model selection and the selection induced bias are important concepts that have received relatively little attention compared to the vast literature on model selection in general.
However, the topic has been discussed for example by \cite{rencher1980}, \cite{ambroise2002}, \cite{reunanen2003}, \cite{varma2006}, and \cite{cawley2010}.
These authors discuss mainly the model selection using cross-validation, but the ideas apply also to other utility estimation methods.
As discussed in Section~\ref{sec:utility_estimation}, cross-validation gives a nearly unbiased estimate of the generalization performance of any given model, but the selection process may overfit when the variance in the utility estimates is high (as depicted in the left plot of Figure~\ref{fig:selectionbias}).
This will be demonstrated empirically in Section~\ref{sec:results}.
The variance in the utility estimate is different for different estimation methods but may generally be considerable for small datasets.
The amount of overfitting in selection increases with the number of models being compared, and may become a problem for example in variable selection where the number of candidate models grows quickly with the number of variables.

\section[Results]{Numerical experiments}
\label{sec:results}

This section compares the methods presented in Section~\ref{sec:methods} in practical variable selection problems.
Section~\ref{sec:models} discusses the used models and Sections~\ref{sec:results_simulated} and \ref{sec:results_real} show illustrative examples using simulated and real world data, respectively.
The reader is encouraged to go through the simulated examples first as they illustrate most of the important concepts with more detailed discussion.
In Section~\ref{sec:modelsize_selection} we then discuss the use of cross-validation for guiding the model size selection and for performance evaluation of the finally selected model.
Finally, Section~\ref{sec:computational_considerations} provides a short note on the computational aspects.

\subsection{Models}
\label{sec:models}

We will consider both regression and binary classification problems.
To reduce the computational burden involved in the experiments, we consider only linear models.
For regression, we apply the standard Gaussian model
\begin{equation}
\begin{split}
	y \given \vc x, \vc w, \sigma^2 &\sim \Normal { \vc w^\tp \vc x, \sigma^2 },  \\
	\vc w \given  \sigma^2, \tau^2 &\sim \Normal {0, \tau^2\sigma^2 \vc I},  \\
	\sigma^2 &\sim \InvGamma{\alpha_\sigma, \beta_\sigma}, \\
	\tau^{2} &\sim \InvGamma{\alpha_\tau, \beta_\tau} \,,
\end{split}
\label{eq:model_reg}
\end{equation}
where $\vc x$ is the $p$-dimensional vector of inputs, $\vc w$ contains the corresponding weights and $\sigma^2$ is the noise variance.
For this model, most of the computations can be obtained analytically because for a given hyperparameter $\tau^2$ the prior is conjugate.
Since it is difficult to come up with a suitable value for the weight regularising variance $\tau^2$, it is given a weakly informative inverse-gamma prior and integrated over numerically.
For the binary classification, we use the probit model 
\begin{equation}
\begin{split}
	y \given \vc x, \vc w &\sim \Ber{ \Phi(\vc w^\tp \vc x) }, \\
	\vc w \given \tau^2 &\sim \Normal{0,\tau^2 \vc I}, \\
	\tau^2 &\sim \InvGamma{\alpha_\tau, \beta_\tau},
\end{split}
\label{eq:model_probit}
\end{equation}
where $\Phi(\cdot)$ denotes the cumulative density of the standard normal distribution.
Again, a weakly informative prior for $\tau^2$ is used.
For this model, we use Markov chain Monte Carlo (MCMC) methods to obtain samples from the posterior of the weights to get the predictions.
For both models~\eqref{eq:model_reg} and~\eqref{eq:model_probit} we include the intercept term by augmenting a constant term in the input vector $\vc x = (1,\, x^1,\dots, x^p)$ and a corresponding term in the weight vector $\vc w = (w^0,w^1,\dots,w^p)$. 
The exact values used for the hyperparameters $\alpha_\tau, \beta_\tau, \alpha_\sigma, \beta_\sigma$ will be given together with the dataset descriptions in Sections~\ref{sec:results_simulated} and \ref{sec:results_real}.

Since we are considering a variable selection problem, the submodels have different number of input variables and therefore different dimensionality for $\vc x$ and $\vc w$.
For notational convenience, the binary vector $\vs \gamma = (\gamma^0,\gamma^1,\dots,\gamma^p)$ denoting which of the variables are included in the model is omitted in the above formulas.
Both in~\eqref{eq:model_reg} and~\eqref{eq:model_probit} the model specification is the same for each submodel $\vs \gamma$, only the dimensionality of $\vc x$ and $\vc w$ change. The reference model $M_*$ is constructed as the BMA~\eqref{eq:bma} from the submodels using the reversible jump MCMC (RJMCMC) \citep{green1995}, which corresponds to a spike-and-slab prior for the full model.
For an additional illustration using a hiearchical shrinkage prior, see Appendix~\ref{app:proj_with_hs}.
For the model space we use the prior
\begin{align}
\begin{split}
	\gamma^j \given \pi &\sim \Ber{\pi}, \quad j=1,\dots,p \,,  \\
	\pi &\sim \Beta{a, b}.
\end{split}
\label{eq:model_prior}
\end{align}
Here parameters $a$ and $b$ adjust the prior beliefs about the number of included variables.
We set $\gamma^0 = 1$, that is, the intercept term $w^0$ is included in all the submodels.
Also for $a$ and $b$, the exact values used will be given together with the dataset descriptions in Sections~\ref{sec:results_simulated} and \ref{sec:results_real}.

\subsection{Simulated data}
\label{sec:results_simulated}

We first introduce a simulated variable selection experiment which illustrates a number of important concepts and the main differences between the different methods. 
The data is distributed as follows
\begin{align}
\begin{split}
	\vc x &\sim \Normal {0, \vc R}, \qquad \vc R \in \mathbb{R}^{p\times p}, \\
	y \given \vc x &\sim \Normal {\vc w^\tp \vc x, \sigma^2}, \hspace{0.1cm} \sigma^2 = 1.
\end{split}
\end{align}
We set the total number of variables to $p=100$.
The variables are divided into groups of 5 variables.
Each variable~$x^j$ has a zero mean and unit variance and is correlated with other variables in the same group with coefficient $\rho$ but uncorrelated with variables in the other groups (the correlation matrix $\vc R$ is block diagonal).
The variables in the first three groups have weights $(w^{1:5},w^{6:10},w^{11:15})=(\xi,\, 0.5\,\xi,\, 0.25\,\xi)$ while the rest of the variables have zero weight.
Thus there are 15 relevant and 85 irrelevant variables in the data.
The constant $\xi$ adjusts the signal-to-noise ratio of the data.
To get comparable results for different levels of correlation $\rho$, we set $\xi$ so that $\sigma^2 / \var y = 0.3$.
For $\rho=0, 0.5, 0.9$ this is satisfied by setting approximately $\xi = 0.59, 0.34, 0.28$, respectively.

\begin{table*}[t]
\abovetopsep=2pt
\caption{  Compared model selection methods for the experiments. MAP and Median models are estimated from the RJMCMC samples, for other methods the searching is done using forward searching (at each step choose the variable that improves the objective function value the most). The methods are discussed in Section~\ref{sec:methods}. }
\label{tab:methods}
\begin{tabular}{ l  p{12.2cm} }
\toprule
Abbreviation & Method \\
\midrule
CV-10 & 10-fold cross-validation optimization~\eqref{eq:kfcv}  \\ 
WAIC & WAIC optimization~\eqref{eq:waic} \\
DIC & DIC optimization~\eqref{eq:dic} \\
L2 & $L^2$-criterion optimization~\eqref{eq:L2} \\
L2-CV & $L^2_\tx{cv}$-criterion optimization~\eqref{eq:L2cv} \\
L2-$k$ & $L^2_k$-criterion optimization with $k=1$~\eqref{eq:L2k}  \\
MAP & Maximum a posteriori model \\
MPP/Median & Sort the variables according to their marginal posterior probabilities (MPP), choose all with probability 0.5 or more (Median)~\eqref{eq:median_model} \\
BMA-ref & Posterior predictive discrepancy minimization from BMA~\eqref{eq:ref_discrepancy}, choose smallest model having 95\% explanatory power~\eqref{eq:explanatory_power} \\
BMA-proj & Projection of BMA to submodels~\eqref{eq:projection_discrepancy_approx}, choose smallest model having 95\% explanatory power~\eqref{eq:explanatory_power} \\
\bottomrule
\end{tabular}
\end{table*}

The experiments were carried out by varying the training set size $n = 100,200,400$ and the correlation coefficient $\rho=0,0.5,0.9$. 
We used the regression model~\eqref{eq:model_reg} with prior parameters $\alpha_\tau = \beta_\tau = \alpha_\sigma = \beta_\sigma = 0.5$.
The posterior inference did not seem to be sensitive to these choices.
As the reference model $M_*$, we used the BMA~\eqref{eq:bma} over the different input combinations with prior $a=1$, $b = 10$ for the number of inputs~\eqref{eq:model_prior}.
For each combination of $(n,\rho)$, we performed the variable selection with each method listed in Table~\ref{tab:methods} for 50 different data realizations.
As a search heuristic, we used the standard forward search, also known as the stepwise regression.
In other words, starting from the empty model, at each step we select the variable that increases the utility estimate (like CV, WAIC, DIC, etc.) the most.
The Median and MAP model where estimated from the RJMCMC samples that were drawn to form the BMA (as discussed in Section~\ref{sec:models}).

\begin{figure*}[t]
	\centering
	\setlength{\figureheight}{0.14\textwidth}
	\setlength{\figurewidth}{0.32\textwidth}
	\pgfplotsset{
	compat=newest,
	title style={yshift=-0em, font=\normalsize},
	legend style={font=\footnotesize},
	x tick label style={font=\footnotesize},
	major tick length={0.07cm},
	}
	\minput[pdf]{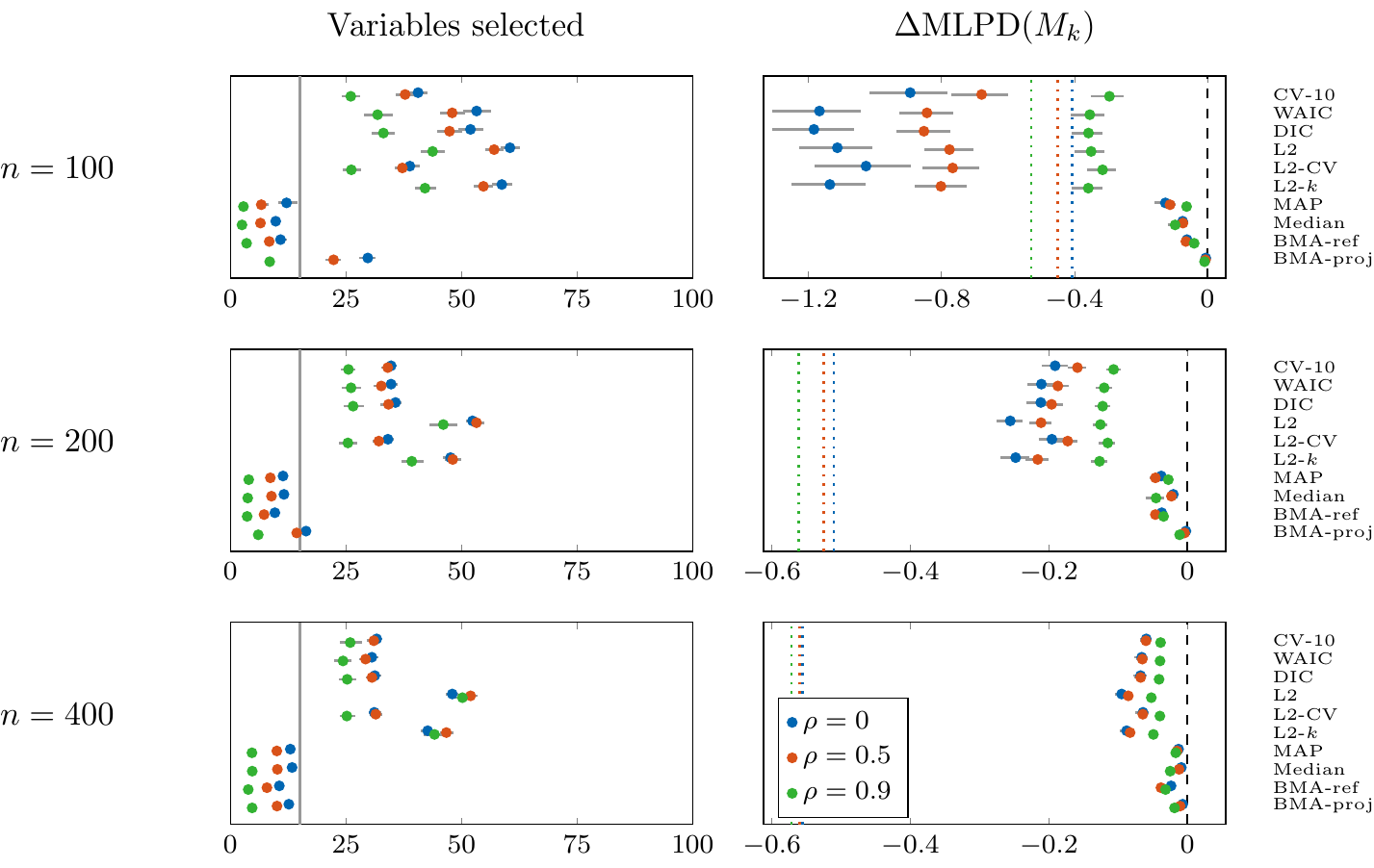}
	\caption{Simulated data: Average forward search paths for some of the selection methods for different training set sizes $n$ when $\rho=0.5$. Red shows the CV utility (10-fold) and black the test utility with respect to the BMA~\eqref{eq:mlpd_diff} after sorting the variables, as a function of number of variables selected averaged over the 50 different data realizations. The difference between these two curves illustrates the selection induced bias. The dotted vertical lines denote the average number of variables chosen with each of the methods (see Table~\ref{tab:methods}).}
   \label{fig:simulated_nsel_mlpd}
\end{figure*}

The found models were then tested on an independent test set of size $\ti n = 1000$.
As a proxy for the generalization utility~\eqref{eq:util_gen}, we use the mean log predictive density (MLPD) on the test set
\begin{align}
	\tx{MLPD}(M) = \frac{1}{\ti n} \sum_{i=1}^{\ti n} \log p(\ti y_i \given \vc{\ti x}_i, \data, M).
\label{eq:mlpd}
\end{align}
To reduce variance over the different data realizations and to better compare the relative performance of the different methods, we report the utilities of the selected submodels $M$ with respect to the gold standard BMA solution $M_*$
\begin{align}
	\Delta \tx{MLPD}(M) = \tx{MLPD}(M) - \tx{MLPD}(M_*).
\label{eq:mlpd_diff}
\end{align}
On this relative scale zero indicates the same predictive performance as the BMA and negative values worse (and positive values better, correspondingly).
A motivation for looking at the relative performance~\eqref{eq:mlpd_diff} is that, as we shall see shortly, the selection typically introduces loss in the predictive accuracy, and we want to assess {\it which of the selection methods are able to find the simplest models with performance close to the BMA}.

\begin{figure*}[t]
	\centering
	\setlength{\figureheight}{0.13\textwidth}
	\setlength{\figurewidth}{0.22\textwidth}
	\pgfplotsset{
	compat=newest,
	title style={yshift=-0em, font=\normalsize},
	ylabel style={rotate=-90, align=left, text width=5em},
	y tick label style={font=\scriptsize},
	x tick label style={font=\scriptsize},
	major tick length={0.07cm},
	axis on top,
	}
	\minput[pdf]{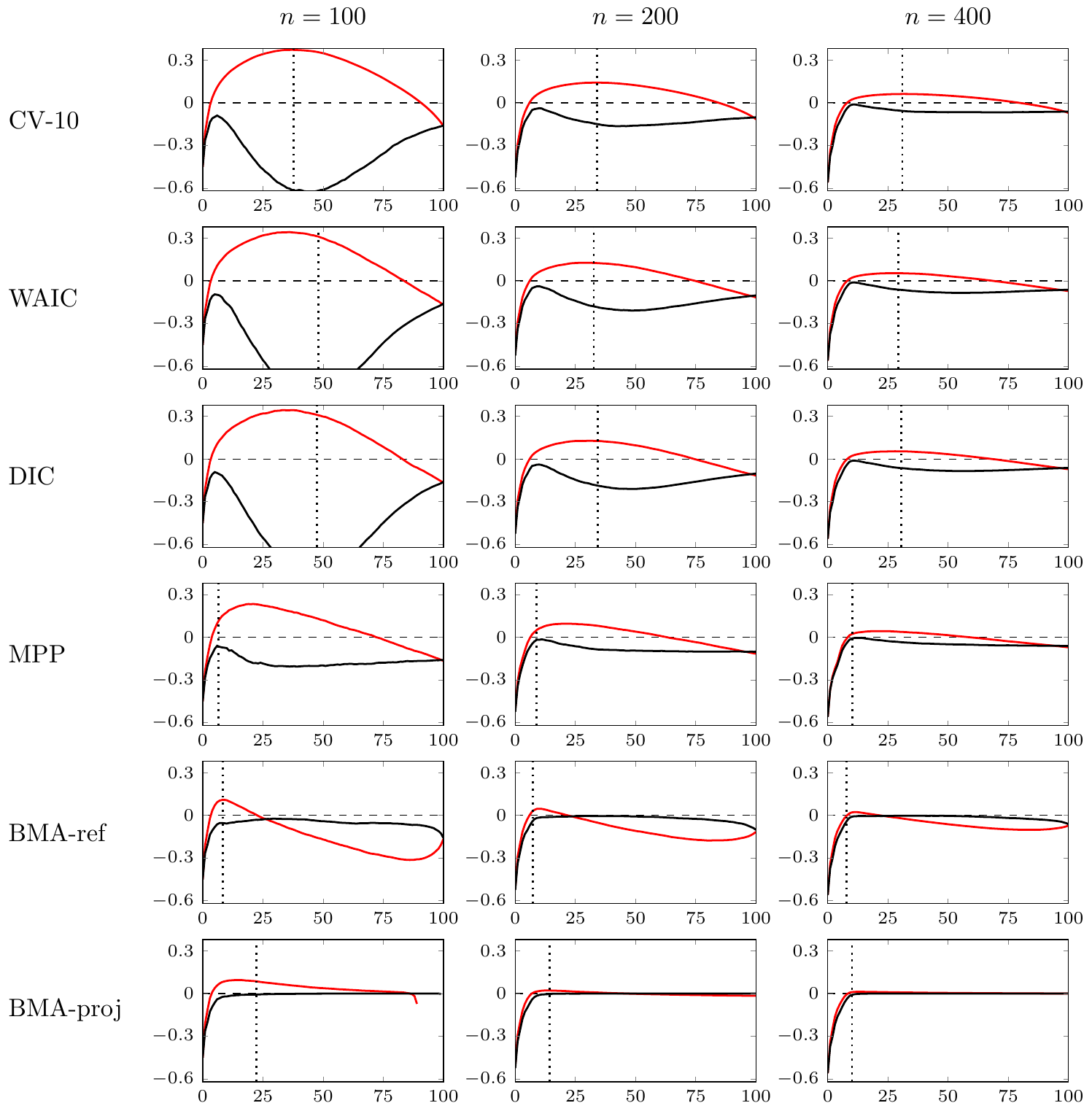}
	\caption{Simulated data: Average forward search paths for some of the selection methods for different training set sizes $n$ when $\rho=0.5$. Red shows the CV utility (10-fold) and black the test utility for the submodels with respect to the BMA~\eqref{eq:mlpd_diff} as a function of number of variables selected averaged over the 50 different data realizations. The difference between these two curves illustrates the selection induced bias. The dotted vertical lines denote the average number of variables chosen with each of the methods (see Table~\ref{tab:methods}).}
   \label{fig:simulated_searchpath}
\end{figure*}

Figure~\ref{fig:simulated_nsel_mlpd} shows the average number of selected variables and the test utilities of the selected models in each data setting with respect to the BMA. 
First, a striking observation is that none of the methods is able to find a model with better predictive performance than the BMA.
From the predictive point of view, model averaging yields generally the best results on expectation, and one should not expect to do better by selection.
This result is in perfect accordance with what is known about the good performance of the BMA \citep{hoeting1999,raftery2003}.
Thus, the primary motivation for selection should be the simplification of the model without substantially compromising the predictive accuracy, rather than trying to improve over the predictions obtained by taking into account the model uncertainty.

\begin{figure*}[t]
	\centering
	\setlength{\figureheight}{0.13\textwidth}
	\setlength{\figurewidth}{0.22\textwidth}
	\pgfplotsset{
	compat=newest,
	title style={yshift=-0em, font=\normalsize},
	ylabel style={rotate=-90, align=left, text width=5em},
	y tick label style={font=\scriptsize},
	x tick label style={font=\scriptsize},
	major tick length={0.07cm},
	axis on top,
	}
	\minput[pdf]{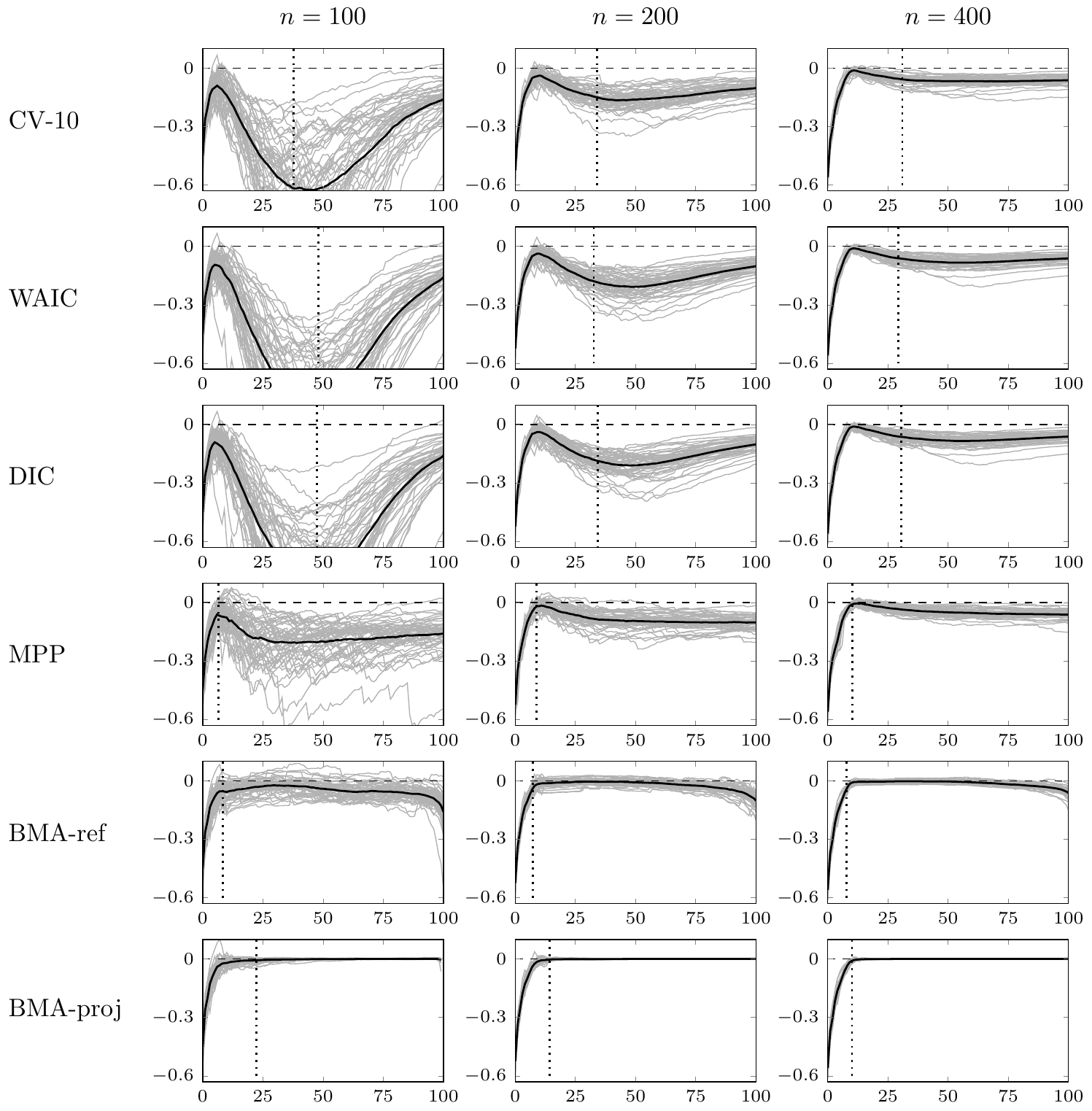}
	\caption{Simulated data: Variability in the predictive performance of the found submodels with respect to the BMA~\eqref{eq:mlpd_diff} along the forward search path as a function of number of variables selected for the same methods as in Figure~\ref{fig:simulated_searchpath} for different training set sizes $n$ when $\rho=0.5$. The grey lines show the test utilities for the different data realizations and the black line denotes the average (the black lines are the same as in Figure~\ref{fig:simulated_searchpath}). The dotted vertical lines denote the average number of variables chosen. }
   \label{fig:simulated_variability}
\end{figure*}

\begin{figure*}[t]
	\centering
	\setlength{\figureheight}{0.13\textwidth}
	\setlength{\figurewidth}{0.22\textwidth}
	\pgfplotsset{
	compat=newest,
	title style={yshift=-0em, font=\normalsize},
	y label style={rotate=-90, align=left, text width=5em},
	y tick label style={font=\scriptsize},
	x tick label style={font=\scriptsize},
	major tick length={0.07cm},
	legend style={overlay},
	axis on top,
	}
	\minput[pdf]{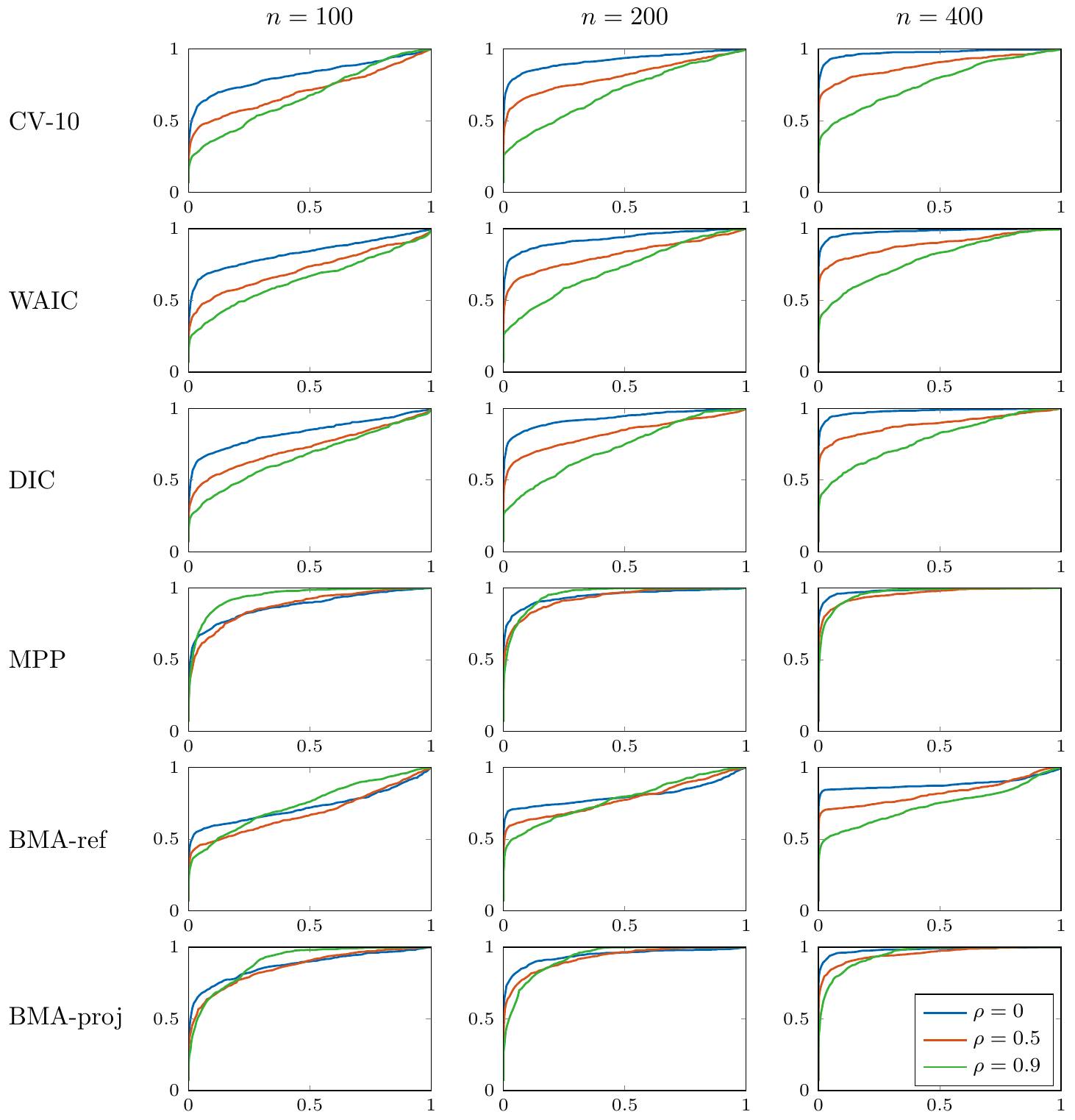}
	\caption{Simulated data: Proportion of relevant (vertical-axis) versus proportion of irrelevant variables chosen (horizontal axis) for the different training set sizes $n$. The data had 100 variables in total with 15 relevant and 85 irrelevant variables, relevant being defined as a variable that was used to generate the output $y$. The colours denote the correlation level between the variables (see the legend). The curves are averaged over the 50 data realizations.}
   \label{fig:simulated_roc}
\end{figure*}

\begin{figure*}[t]
	\centering
	\setlength{\figureheight}{0.32\textwidth}
	\setlength{\figurewidth}{0.82\textwidth}
	\pgfplotsset{
	compat=newest,
	title style={yshift=-0em, font=\normalsize},
	ylabel style={rotate=-90, align=left, text width=5em},
	y tick label style={font=\scriptsize},
	x tick label style={font=\scriptsize},
	major tick length={0.07cm},
	axis on top,
	}
	\minput[pdf]{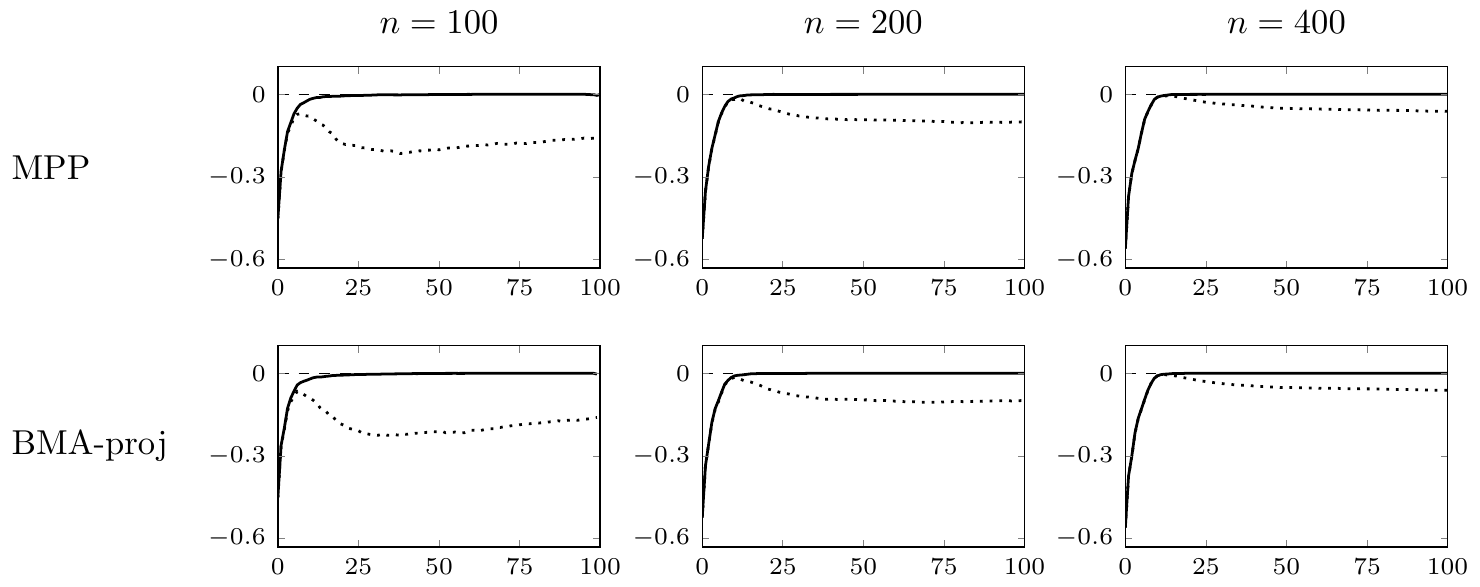}
	\caption{Simulated data: The average test utility with respect to the BMA~\eqref{eq:mlpd_diff} as a function of number of variables selected when the submodel parameters are learned by projection from the BMA (solid line) and by standard fitting to the data (dotted line). The projection improves the performance of the submodels regardless of whether the variables are sorted by their marginal posterior probabilities (top row) or by a forward search minimizing the discrepancy to the BMA (bottom row).}
   \label{fig:proj_vs_noproj}
\end{figure*}

Second, for the smallest dataset size many of the methods perform poorly and choose models with bad predictive performance, comparable or even worse than the model with no variables at all (the dotted lines).
This holds for CV-10, WAIC, DIC, L2, L2-CV, and L2-$k$, and the conclusion covers all the levels of correlation between the variables (blue, red and green circles), albeit the high dependency between the variables somewhat improves the results.
The observed behaviour is due to overfitting in the selection process (as we will show below).
Due to scarce data, the high variance in the utility estimates leads to selecting overfitted models as discussed in Section~\ref{sec:selection_bias}.
These methods perform reasonably only for the largest dataset size $n=400$.
MAP, Median, BMA-ref, and BMA-proj perform significantly better, choosing smaller models with predictive ability closer to the BMA.
A closer inspection reveals that out of these four, BMA-proj performs best in terms of the predictive ability especially for the smallest dataset size $n=100$, but the better predictive accuracy is partially due to selecting more variables than the other three methods. 
Note also that for BMA-proj the predictions are computed using the projected samples (Eq.~\eqref{eq:projection_pred}), whereas for all the other methods the parameter estimation is done by fitting the submodels to data.
Later in this section we will show that the parameter estimation using the projection can play a considerable role in achieving improved predictive accuracy for the submodels, and the good performance of BMA-proj is not simply due to superior ordering of the variables, in fact MPP may perform even better in this aspect (see discussion related to Figures~\ref{fig:simulated_roc}~and~\ref{fig:proj_vs_noproj}).

To get more insight to the problem, let us examine more closely how the predictive performance of the submodels change when variables are selected.
Figure~\ref{fig:simulated_searchpath} shows the CV and test utilities {\it after} sorting the variables, as a function of the number of variables selected along the forward search path when $\rho=0.5$.
The CV-utility (10-fold) is computed within the data used for selection ($n$ points), and the test utility on independent data (note that computing the CV-curve for BMA-proj requires cross-validating the BMA and performing the projection for the submodels separately for each fold).
The search paths for CV-10 (top row) demonstrate the overfitting in model selection; starting from the empty model and adding variables one at a time one finds models that have high CV utility but much worse test utility. 
In other words, the performance of the models at the search path is dramatically overestimated and the gap between the two curves denotes the selection induced bias.
Yet in other words, {\it after} selection (sorting the variables) the CV utility is an optimistic estimate for the selected models.
Note, however, that for the empty model and the model with all the variables the CV utility and the test utility are on average almost the same because these models do not involve any selection.
The overfitting in the selection process decreases when the size of the training set grows because the variance of the error term in decomposition~\eqref{eq:util_decomp} becomes smaller, but the effect is still visible for $n=400$. 
The behaviour is very similar also for WAIC, DIC, L2, L2-CV and L2-$k$ (the results for the last three are left out to save space).

Ordering the variables according to their marginal posterior probabilities (MPP) and selecting the Median model works much better than CV-10, leading to selection of smaller models with good predictive performance.
However, even better results are obtained by using the reference model approach, especially the projection (BMA-proj).
The results clearly show that the projection approach is much less vulnerable to the overfitting than CV, WAIC and DIC, even though the CV utility is still a biased estimate of the true predictive ability for the chosen models.
Even for the smallest dataset size, the projection is able to find models with predictive ability very close to the BMA with about 10--15 variables on average.
Moreover, the projection has the inherent advantage over the other methods performing reasonably well (like MPP/Median) that when more variables are added, the submodels get ever closer to the reference model (BMA), thus avoiding the dip in the predictive accuracy apparent with the other methods around 10 variables.
This is simply because the submodels are constructed to be similar than the model averaging solution which yields the best results (this point will be further discussed below, see discussion related to Figure~\ref{fig:proj_vs_noproj}).

Figure~\ref{fig:simulated_variability}  shows the variability in the performance of the selected models for the same selection methods as in Figure~\ref{fig:simulated_searchpath}.
The grey lines denote the test utilities for the selected models as a function of number of selected variables for different data realizations and the black line denotes the average (same as in Figure~\ref{fig:simulated_searchpath}). 
For small training set sizes the variability in the predictive performance of the selected submodels is very high for CV-10, WAIC, DIC, and MPP.
The reference model approach, especially the projection, reduces the variability substantially finding sparse submodels with predictive performance close to the BMA in all the data realizations.
This is another property that makes the projection approach appealing.
Figure~\ref{fig:simulated_variability} further emphasizes how difficult it is to improve over the BMA in predictive terms; most of the time the model averaging yields better predictions than any of the found submodels.
Moreover, even when better submodels are found (the cases where the grey lines exceed the zero level), the difference in the predictive performance is relatively small.

Although our main focus is on the predictive ability of the chosen models, we also studied how the different methods are able to choose the truly relevant variables over the irrelevant ones.
Here by ``relevant'' we mean those 15 variables that were used to generate the output $y$ even though it might not be completely clear how the relevance should be defined when there are correlations between the variables (for example, should we select one or both of two correlating variables which both correlate with the output).
Figure~\ref{fig:simulated_roc} shows the proportion of relevant variables chosen (vertical axis) versus proportion of irrelevant variables chosen (horizontal axis) on average (the larger the area under the curve, the better).
In this aspect, ordering the variables according to their marginal probabilities seems to work best, slightly better than the projection.
The other methods seem to perform somewhat worse.
Interestingly, although the projection does not necessarily order the variables any better than the marginal posterior probability order, the predictive ability of the projected submodels is on average better and varies less as Figure~\ref{fig:simulated_variability} demonstrates.

To explain this behaviour, we did one more experiment and studied the difference of learning the submodel parameters by the projection from the BMA compared to fitting the submodels to the data.
We performed this analysis both when the selection was done by the marginal probabilities or by forward search minimizing the discrepancy to the BMA, see Figure~\ref{fig:proj_vs_noproj}.
The results show that constructing the submodels by projection improves the results regardless of which of the two methods is used to sort the variables, and in this example, there seems to be little difference in the final results as long as projection is used to construct the submodels.

This effect can be explained by transimission of information from the reference model. 
Recall that the BMA corresponds to setting the sparsifying spike-and-slab prior for the full model.
Because the prior information is transmitted also to the submodels in the projection, it is natural that the submodels benefit from this (compared to the Gaussian prior in~\eqref{eq:model_reg}) especially when some irrelevant variables have been included.
Furthermore, the noise level is also learned from the full model (see Eq.~\eqref{eq:proj_noise}), which reduces overfitting of the submodels as the full model best represents the uncertainties related to the problem and best captures the correct the noise level.
More detailed analysis of these effects would be useful, but we do not focus on it more in this paper and leave it to the future research.

To summarize, it clearly appears that the full model averaging solution produces best predictive results, and the projection appears the most robust method for simplifying the full model without losing much predictive accuracy.
However, the number of variables actually selected depends on the arbitrary 95\% explanatory power rule, and although it seems work quite well for the examples above, it does not always lead to optimal results (see the real world examples in the next section).
We discuss this problem and a possible solution further in Section~\ref{sec:modelsize_selection}.

\subsection{Real world datasets}
\label{sec:results_real}

We also studied the performance of the different methods on several real world datasets.
Five publicly available\footnote{The first three datasets are available at the UCI Machine Learning repository \url{https://archive.ics.uci.edu/ml/index.html}. \\ Ovarian cancer dataset can be found at \url{http://www.dcs.gla.ac.uk/~srogers/lpd/lpd.html} and the Colon cancer data at \url{http://genomics-pubs.princeton.edu/oncology/affydata/index.html}.} datasets were used and they are summarized in Table~\ref{tab:datasets}.
One of the datasets deals with regression and the rest with binary classification.
As a preprocessing, we normalized all the input variables to have zero mean and unit variance.
For the Crime dataset we also log-normalized the original non-negative target variable (crimes per population) to get a real-valued and more Gaussian output.
From this dataset we also removed some input variables and observations with missing values (the given $p$ and $n$ in Table~\ref{tab:datasets} are after removing the missing values).
We will not discuss the datasets in detail but refer to the sources for more information.

\begin{figure*}%
	\centering
	\setlength{\figureheight}{0.25\textwidth}
	\setlength{\figurewidth}{1.0\textwidth}
	\pgfplotsset{
	compat=newest,
	scaled ticks=false,
	title style={font=\normalsize},
	y tick label style={font=\tiny, /pgf/number format/fixed},
	x tick label style={font=\tiny},
	x label style={font=\scriptsize},
	major tick length={0.07cm},
	legend style={font=\tiny},
	}
	\minput[pdf]{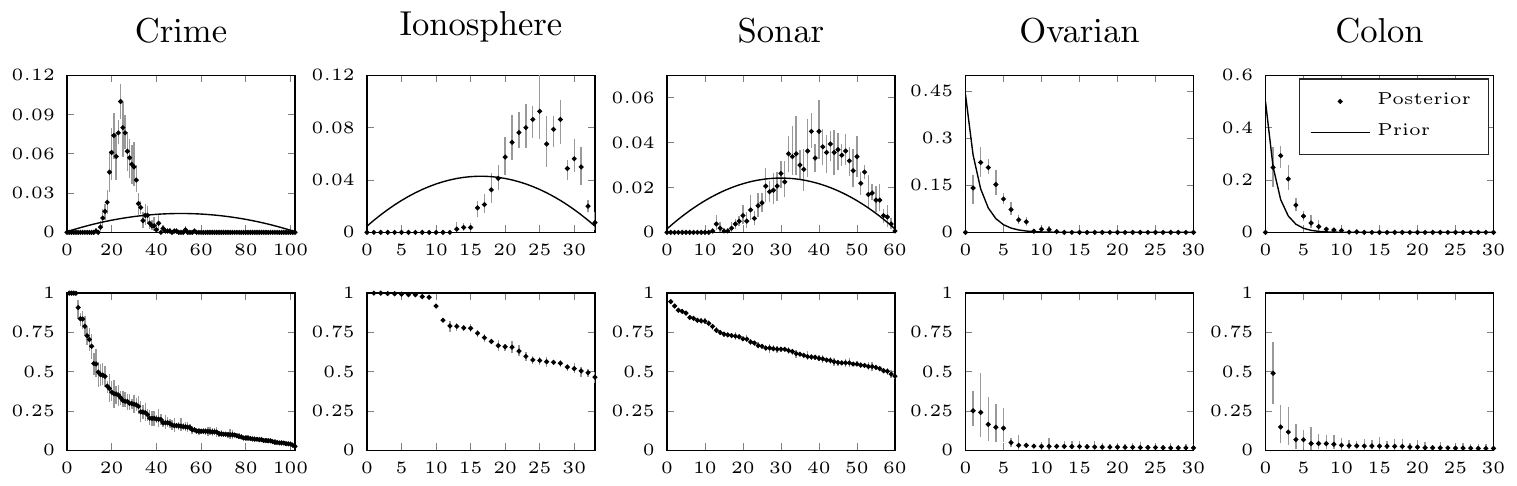}
	\caption{Real datasets: Prior and posterior probabilities for the different number of variables (top row) and marginal posterior probabilities for the different variables sorted from the most probable to the least probable (bottom row). The posterior probabilities are given with 95\% credible intervals estimated from the variability between different RJMCMC chains. The results are calculated using the full datasets (not leaving any data out for testing). For Ovarian and Colon datasets the plots are truncated at 30 variables.}
  \label{fig:modelspace}
\end{figure*}

For the regression problem we applied the Gaussian regression model~\eqref{eq:model_reg} and for the classification problems the probit model~\eqref{eq:model_probit}. 
The prior parameters in each case are listed in Table~\ref{tab:datasets}.
For all the problems we used relatively uninformative priors for the input weights (and measurement noise).
As the reference model, we again used the BMA solution estimated using reversible jump MCMC.
For the first three problems (Crime, Ionosphere, Sonar) we used a very uninformative prior for the number of input variables (i.e., $a=b=2$) because there was basically no prior information about the sparsity level.
For the last two datasets (Ovarian and Colon) for which $p \gg n$ we had to use priors that favor models with only a few variables to avoid overfitting.
Figure~\ref{fig:modelspace} shows the estimated posterior probabilities for different number of variables (top row) and marginal posterior probabilities for the different inputs (bottom row) for all the datasets.
Although these kind of plots may give some idea about the variable relevancies, it is still often difficult to decide which variables should be included in the model and what would be the effect on the predictive performance.

\begin{table*}%
\centering
\abovetopsep=2pt
\caption{Summary of the real world datasets and used priors. $p$ denotes the total number of input variables and $n$ is the number of instances in the dataset (after removing the instances with missing values). The classification problems deal all with a binary output variable.}
\label{tab:datasets}
\begin{tabular}{ llccp{7cm} }
\toprule
Dataset & Type & $p$ & $n$ & Prior parameters\\ 
\midrule
Crime & Regression & 102 & 1992 & ${\alpha_\tau = \beta_\tau = 0.5}$, ${\alpha_\sigma = \beta_\sigma = 0.5}$, ${a = b = 2}$  \\ 
Ionosphere & Classification & 33 & 351 & ${\alpha_\tau = \beta_\tau = 0.5},\, {a = b = 2}$ \\ 
Sonar & Classification & 60 & 208 & ${\alpha_\tau = \beta_\tau = 0.5},\, {a = b = 2}$  \\ 
Ovarian cancer & Classification & 1536 & 54 & ${\alpha_\tau = \beta_\tau = 2},\, {a = 1,\, b = 1200}$  \\
Colon cancer & Classification & 2000 & 62 & ${\alpha_\tau = \beta_\tau = 2},\, {a = 1,\, b = 2000}$ \\
\bottomrule
\end{tabular}
\end{table*}

We then performed the variable selection using the methods in Table~\ref{tab:methods} except the ones based on the squared error (L2, L2-CV, L2-$k$) were not used for the classification problems.
For Ovarian and Colon datasets, due to large number of variables, we also replaced the 10-fold-CV by the importance sampling LOO-CV (IS-LOO-CV) to reduce the computation time.
For these two datasets we also performed the forward searching only up to 10 variables.
To estimate the predictive ability of the chosen models, we repeated the selection several times each time leaving part of the data out and then measuring the out-of-sample performance using these observations.
The Crime dataset was sufficiently large ($n=1992$) to be splitted into training and test sets.
We repeated the selection for 50 random splits each time using $n=100,200,400$ points for training and the rest for testing.
This also allowed us to study the effect of the training set size.
For Ionosphere and Sonar we used 10-fold cross-validation, that is, the selection was performed 10 times each time using 9/10 of the data and estimating the out-of-sample performance with the remaining 1/10 of the data.
For Ovarian and Colon datasets, due to few observations, we used leave-one-out cross-validation for performance evaluation (the selection was performed $n$ times each time leaving one point out for testing).
Again, we report the results as the mean log predictive density on the independent data with respect to the BMA~\eqref{eq:mlpd_diff}.

\begin{figure*}[tp]
	\centering
	\setlength{\figureheight}{0.13\textwidth}
	\setlength{\figurewidth}{0.32\textwidth}
	\pgfplotsset{
	compat=newest,
	title style={yshift=-0em, font=\normalsize},
	tick label style={/pgf/number format/fixed},
	y label style={rotate=-90, align=left, text width=8em},
	x tick label style={font=\scriptsize, /pgf/number format/fixed},
	major tick length={0.07cm},
	}
	\minput[pdf]{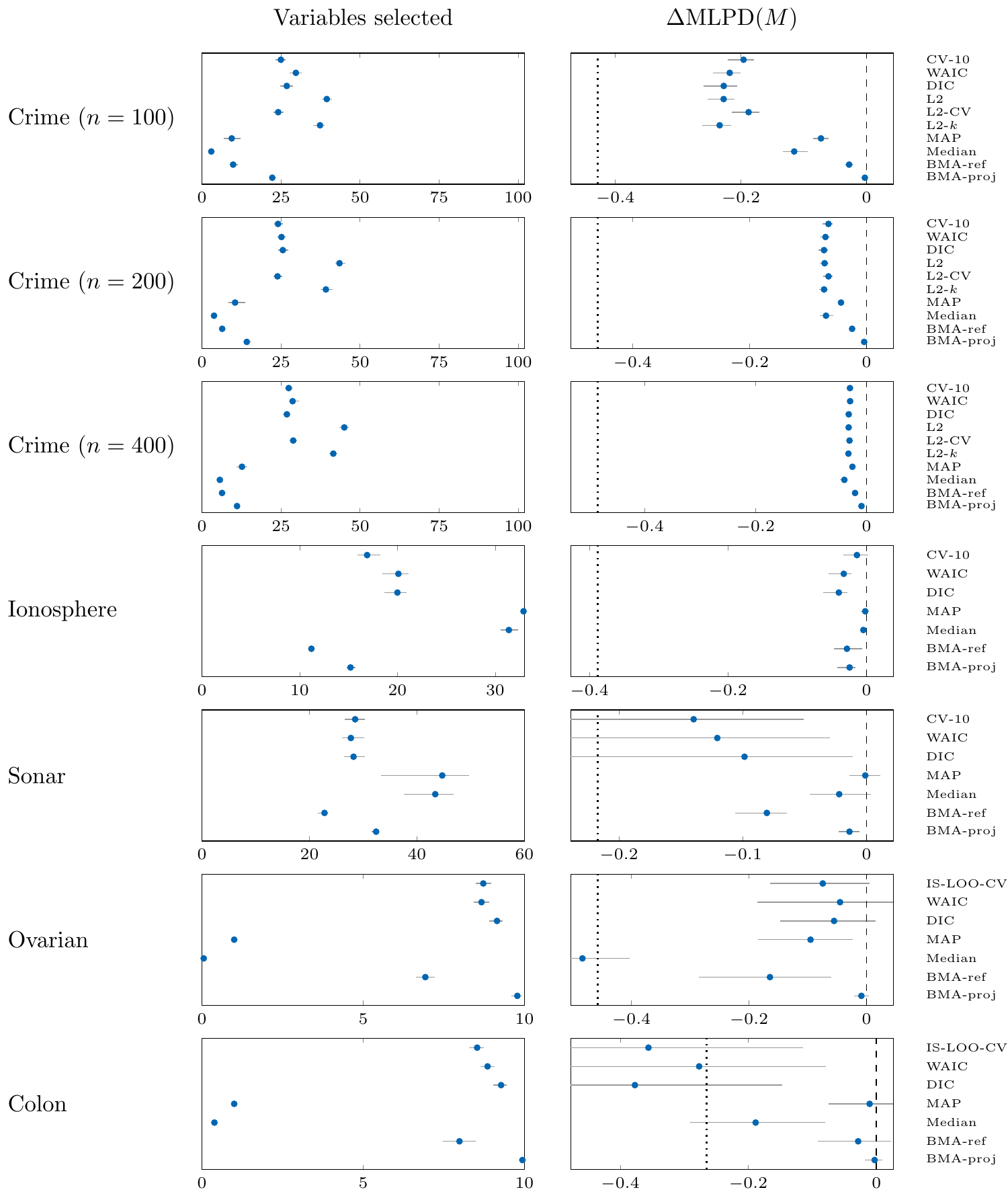}
	\caption{Real datasets: The number of selected variables (left column) and the estimated out-of-sample utilities of the selected models (right column) on average and with 95\% credible intervals for the different datasets. The out-of-sample utilities are estimated using independent data not used for selection (see text) and are shown with respect to the BMA~\eqref{eq:mlpd_diff}. The dotted line denotes the performance of the empty model (the intercept term only). For Ovarian and Colon datasets the searching was performed only up to 10 variables although both of these datasets contain many more variables.}
   \label{fig:real_nsel_mlpd}
\end{figure*}

\begin{figure*}[tp]
	\centering
	\setlength{\figureheight}{0.11\textwidth}
	\setlength{\figurewidth}{0.15\textwidth}
	\pgfplotsset{
	compat=newest,
	title style={yshift=-0em, font=\normalsize},
	y label style={rotate=-90, align=left, text width=5.5em},
	y tick label style={font=\scriptsize},
	x tick label style={font=\scriptsize},
	major tick length={0.07cm},
	}
	\minput[pdf]{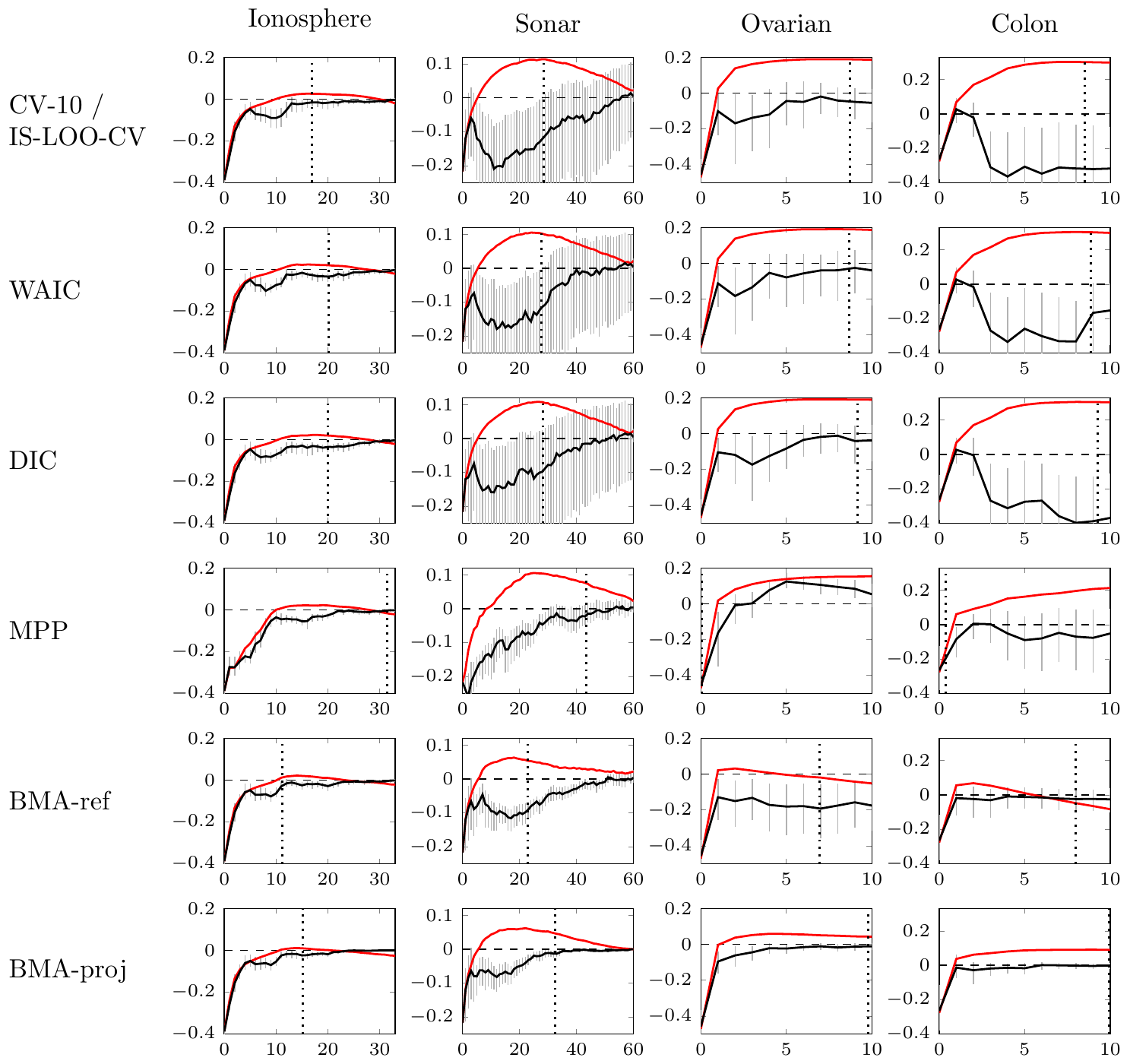}
	\caption{Classification datasets: CV (red) and out-of-sample (black) utilities on average for the selected submodels with respect to the BMA~\eqref{eq:mlpd_diff} along the forward search path as a function of number of variables selected. CV utilities (10-fold) are computed within the same data used for selection and the out-of-sample utilities are estimated on hold-out samples not used for selection (see text) and are given with 95\% credible intervals. The dotted vertical lines denote the average number of variables chosen. CV optimization (top row) is carried out using 10-fold-CV for Ionosphere and Sonar, and IS-LOO-CV for Ovarian and Colon. }
   \label{fig:searchpath_classific}
\end{figure*}
\begin{figure*}
	\centering
	\setlength{\figureheight}{0.13\textwidth}
	\setlength{\figurewidth}{0.22\textwidth}
	\pgfplotsset{
	compat=newest,
	title style={yshift=-0em, font=\normalsize}, 
	y label style={rotate=-90, align=left, text width=5em},
	y tick label style={font=\scriptsize},
	x tick label style={font=\scriptsize},
	major tick length={0.07cm},
	axis on top,
	}
	\minput[pdf]{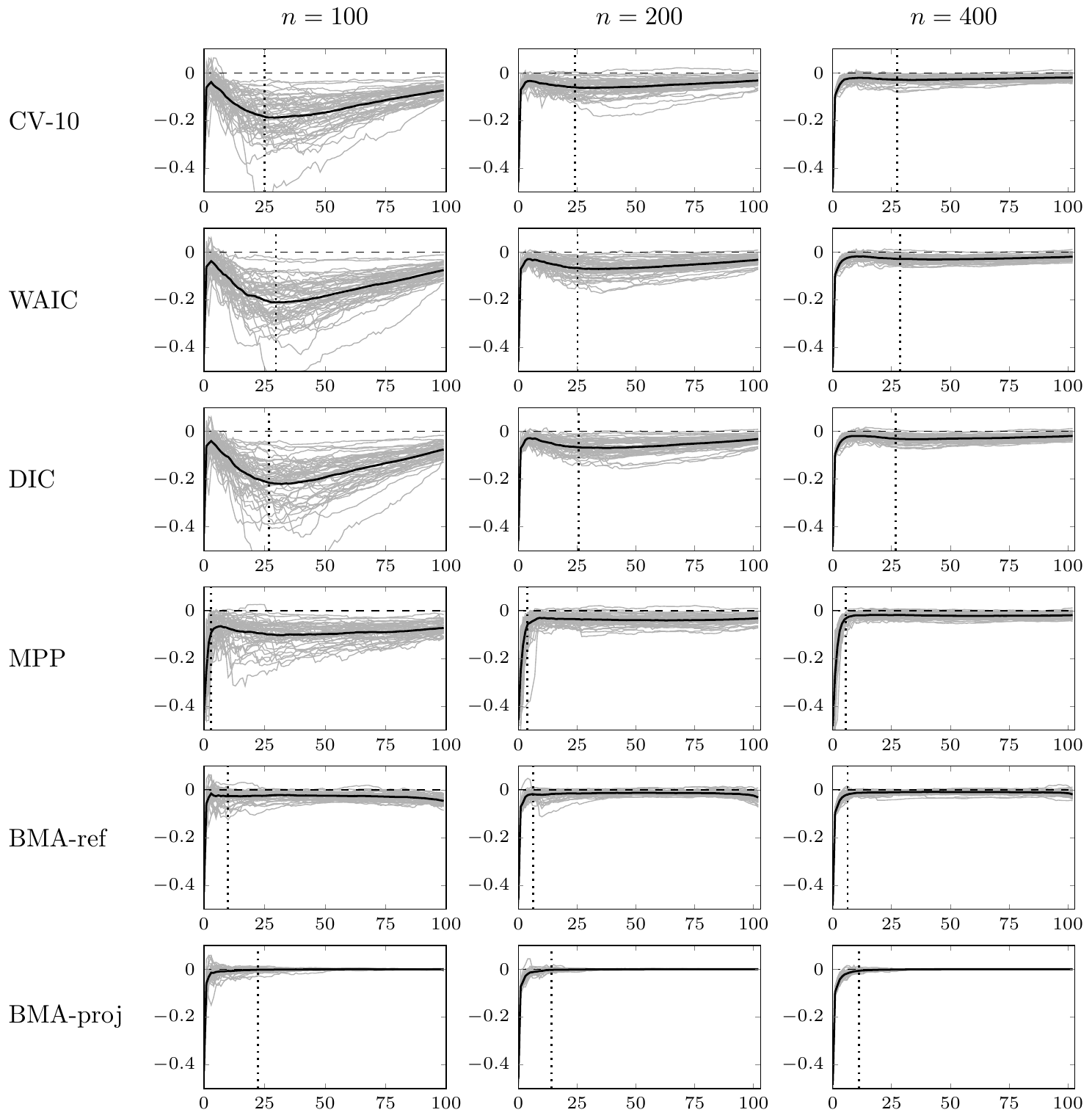} 
	\caption{Crime dataset: Variability in the test utility of the selected submodels with respect to the BMA~\eqref{eq:mlpd_diff} along the forward search path as a function of number of variables selected. The selection is performed using $n=100,200,400$ points and the test utility is computed using the remaining data.  The grey lines show the test utilities for the 50 different splits into training and test sets and the black line denotes the average. The dotted vertical lines denote the average number of variables chosen.}
   \label{fig:searchpath_crime}
\end{figure*}

Figure~\ref{fig:real_nsel_mlpd} summarizes the results.
The left column shows the average number of variables selected and the right column the estimated out-of-sample utilities for the chosen models (out-of-sample utilities are estimated using hold-out samples not used for selection as explained above).
The results are qualitatively very similar to those obtained for the simulated experiments (Sec. \ref{sec:results_simulated}).
Again we conclude that the BMA solution gives better predictions than any of the selection methods when measured on independent data.
Moreover, the results demonstrate again that model selection using CV, WAIC, DIC, L2, L2-CV, or L2-$k$ is liable to overfitting especially when the dataset is small compared to the number of variables.
Overall, MAP and Median models tend to perform better but show non-desirable performance on some of the datasets.
Especially the Median model performs badly for Ovarian and Colon datasets where almost all the variables have marginal posterior probability less than 0.5 (depending on the split into training and validation sets).
The projection (BMA-proj) shows the most robust performance choosing models with predictive ability close to the BMA for all the datasets.

Figure~\ref{fig:searchpath_classific} shows the CV (red) and out-of-sample (black) utilities for the chosen models as a function of number of chosen variables for the classification problems. 
CV-utilities (10-fold) are computed after sorting the variables within the same data used for selection, and the out-of-sample utilities are estimated on hold-out samples not used for selection as explained earlier.
This figure is analogous to Figure~\ref{fig:simulated_searchpath} showing the magnitude of the selection induced bias (the difference between the red and black lines).
Especially for the last three datasets (Sonar, Ovarian, Colon) the selection induced bias is considerable for all the methods, which emphasizes the importance of validating the variable searching process in order to avoid bias in performance evaluation for the found models.
Overall, the projection (BMA-proj) appears to find models with best out-of-sample accuracy for a given model complexity, albeit for Ovarian dataset choosing about five most probable inputs (MPP) would perform even better.
Moreover, the uncertainty in the out-of-sample performance for a given number of variables is also the smallest for the projection over all the datasets.

The same applies to the Crime dataset, see Figure~\ref{fig:searchpath_crime}.
For any given number of variables the projection is able to find models with predictive ability closest to the BMA and also with the least variability, the difference to the other methods being the largest when the dataset size is small.
For Crime data, some additional results using hiearchical shrinkage prior for the full model are presented in Appendix~\ref{app:crime_with_hs}.

\FloatBarrier
\clearpage

\subsection{On choosing the final model size}
\label{sec:modelsize_selection}

Although the search paths for the projection method (BMA-proj) seem overall better than for the other methods (Figures~\ref{fig:simulated_searchpath}, \ref{fig:simulated_variability}, \ref{fig:searchpath_classific} and \ref{fig:searchpath_crime}), the results also demonstrate difficulty in deciding the final model size; for instance, for Ionosphere and Sonar datasets the somewhat arbitrary 95\% explanatory power rule chooses rather too few variables, but for Ovarian and Colon unnecessarily many variables (the out-of-sample utility close to the BMA can be obtained with fewer variables).
The same applies for the Crime dataset with the smallest number of training points ($n=100$).
As discussed in Section~\ref{sec:reference_methods}, a natural idea would be to decide the final model size based on the estimated out-of-sample utility (the black lines in Figures~\ref{fig:searchpath_classific} and \ref{fig:searchpath_crime}) which can be done by cross-validation outside the searching process.
This opens up the question, does this induce a significant amount of bias in the utility estimate for the finally selected model?

\begin{figure*}[t]
	\centering
	\setlength{\figureheight}{0.14\textwidth}
	\setlength{\figurewidth}{0.23\textwidth}
	\pgfplotsset{
	compat=newest,
	title style={yshift=-0.5em, font=\normalsize},
	y tick label style={font=\footnotesize},
	x tick label style={font=\footnotesize},
	x label style={font=\footnotesize},
	major tick length={0.07cm},
	legend style={at={(1,1)}, anchor=north west, xshift=1cm, font=\footnotesize},
	}
	\minput[pdf]{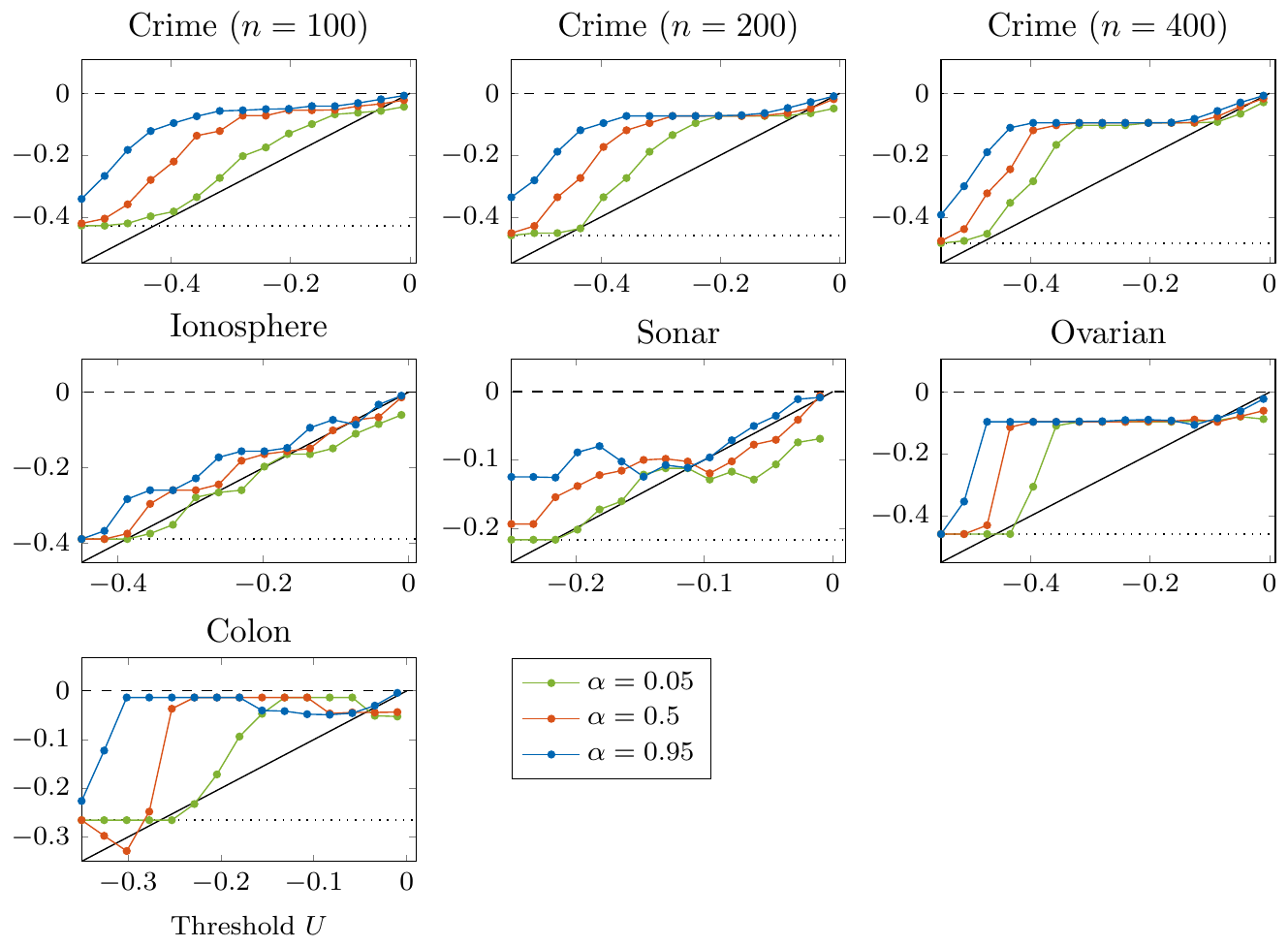}
	\caption{Real datasets: Vertical axis shows the final expected utility on independent data with respect to the BMA~\eqref{eq:mlpd_diff} for the selected submodels when the searching is done using the projection (BMA-proj) selecting the smallest number of variables $m$ satisfying $\pr{\Delta \tx{MLPD}(m) \ge U} \ge \alpha$, where $\Delta \tx{MLPD}(m)$ denotes the estimated out-of-sample utility for $m$ variables estimated using the CV (10-fold) outside the searching process (same as the black lines in Figure~\ref{fig:searchpath_classific}). The final utility is estimated using another layer of validation (see text). The dotted line denotes the utility for the empty model. When $\alpha=0.95$, the final utility remains equal or larger than $U$ (the dots stay above the diagonal line) indicating that the applied selection rule does not induce bias in the performance evaluation for the finally selected model.}
   \label{fig:real_dblcv}
\end{figure*}

To assess this question, we performed one more experiment on the real world datasets by adding another layer of cross-validation to assess the performance of the finally selected models on independent data.
In other words, the variable searching was performed using the projection, the inner layer of cross-validation (10-fold) was used to decide the model size and the outer layer to measure the performance of the finally selected models (10-fold for Ionosphere and Sonar, LOO-CV for Ovarian and Colon, and hold-out with different training set sizes for Crime).
As the rule for deciding the model size, we selected the smallest number of variables satisfying 
\begin{align*}
\pr{\Delta \tx{MLPD}(m) \ge U} \ge \alpha
\end{align*}
for different thresholds $U$ and $\alpha$.
Here $\Delta \tx{MLPD}(m)$ denotes the estimated out-of-sample utility for $m$ variables in the inner validation. 
This probability is the same as~\eqref{eq:pr_u_diff}, the inequality is merely organized differently.
Here $U$ denotes how much one is willing to sacrifice the predictive accuracy in order to reduce the number of variables, and $\alpha$ denotes how certain we want to be about not going below $U$.
We estimate the above probability using the Bayesian bootstrap.

Figure~\ref{fig:real_dblcv} shows the final expected utility on independent data for different values of $U$ and $\alpha$ for the different datasets.
The results show that for a large enough credible level $\alpha=0.95$, the applied selection rule appears to be safe in the sense that the final expected utility remains always above the level of $U$ (the dots stay above the diagonal line).
This means that there is no substantial amount of selection induced bias at this stage, and the second level of validation is not necessarily needed.
However, this does not always hold for smaller values of $\alpha$ ($\alpha=0.5$ and $\alpha=0.05$).

We can make these results more intuitive by looking at an example.
Consider the projection search path for the Sonar dataset in Figure~\ref{fig:searchpath_classific} (bottom row, second column).
Assume now that $U = -0.01$, meaning that we are willing to lose $0.01$ in the MLPD compared to the BMA, which is about 5\% of the difference between the BMA and the empty model.
Since the grey bars denote the central 95\% credible intervals, setting $\alpha=0.975$ would correspond to choosing the smallest model for which the {\it lower} limit of the grey bar falls above $U=-0.01$  (because then the performance of the submodel is greater than this with probability 0.975).
This would lead to choosing about 35 variables, and we could be quite confident that the final expected utility would be within 0.01 from the BMA. 
On the other hand, setting $\alpha=0.025$ corresponds to choosing the smallest model for which the {\it upper} limit of the grey bar falls above $U=-0.01$ (because then the performance of the submodel is greater than this with probability 0.025).
This would lead to choosing only 3 variables, but in this case we cannot expect confidently that the final expected utility would be within 0.01 from the BMA, and in fact it would be lower than this with high probability.

Following the example above, one can get an idea of the effect of $\alpha$ and $U$ to the final model size and the final expected utility.
Generally speaking, $U$ determines how much we are willing to compromise the predictive accuracy and $\alpha$ determines how condident we want to be about the final utility estimate.
It is application specific whether it is more important to have high predictive accuracy or to reduce the number of variables at the cost of losing predictive accuracy, but a reasonable recommendation might be to use $\alpha \ge 0.95$ and $U$ to be 5\% of the difference between the reference model and the empty model.
Based on Figure~\ref{fig:real_dblcv}, this combination would appear to give predictive accuracy close to the BMA, and based on Figure~\ref{fig:searchpath_classific} yield quite effective reduction in the number of variables (choosing about 20 variables for Ionosphere, 35 for Sonar, and 5--10 variables for Ovarian and Colon by visual inspection).

As a final remark, one might wonder why the cross-validation works well if used only to decide the model size but poorly if used directly to optimize the variable combination (as depicted e.g. in Figure~\ref{fig:simulated_searchpath})?  As discussed in Section~\ref{sec:selection_bias}, the amount of overfitting in the selection and the consequent selection bias depends on the variance of the utility estimate for a given model (over the data realizations) and the number of candidate models being compared.
For the latter reason, the overfitting is considerably smaller when cross-validation is used only to decide the model size by comparing only $p+1$ models (given the variable ordering), in contrast to selecting variable combination in the forward search phase among $O(p^2)$ models.
Moreover, the utilities of two consecutive models at the search path are likely to be highly correlated, which further reduces the freedom of selection and therefore the overfitting at this point.
It is for this same reason why cross-validation may yield reasonable results when used to choose a single hyperparameter among a small set of values, in which case essentially only a few different models are being compared.

Based on the results presented in this section, despite the increased computational effort, we believe the use of cross-validation on top of the variable searching is highly advisable both for choosing the final model size and giving a nearly unbiased estimate of the out-of-sample performance for the selected model.
We emphasize the importance of this regardless of the method used for searching the variables, but generally we recommend using the projection given its overall superior performance in our experiments.

\subsection{Computational considerations}
\label{sec:computational_considerations}

We conclude with a remark on the computational aspects.
The results in Sections \ref{sec:results_simulated} and \ref{sec:results_real} emphasized that from the predictive point of view, the model averaging over the different submodels yields often better results than selection of any single model.
Forming the model averaging solution over the variable combinations may be computationally challenging, but is quite feasible up to problems with a few thousand variables or less with, for instance, a straighforward implementation of the RJMCMC algorithm with simple proposal distributions.
Scalability up to a million variables can be obtained with more sophisticated and efficient proposals \citep{peltola2012b}.
The computations could also be sped up by using a hierarchical shrinkage prior such as the horseshoe (see Appendix~\ref{app:hier_shrinkage}) for the regression weights in the full model, instead of the spike-and-slab (which is equivalent to the model averaging over the variable combinations).

After forming the full reference model (either the BMA or using some alternative prior), the subsequent computations needed for the actual selection are typically less burdensome.
Computing the MAP and the Median models from the MCMC samples from the model space is easy, but these cannot be computed with alternative priors.
Constructing the submodels by projection and performing a forward search through the variable combinations is somewhat more laborous, but takes still usually considerably less time than forming the reference model.
The projection approach can also be used with any prior for the full model, as the posterior samples are all that is needed (see Appendix~\ref{app:proj_with_hs}).\sloppy

The methods that do not rely on the construction of the full reference model (CV, WAIC, DIC, and the $L^2$-variants) are typically computationally easier as they avoid the burden of forming the reference model in the first place.
However, if one has to go through a lot of models in the search, the procedure is fast only if the fitting of the submodels is fast.
This is the case for instance for the Gaussian linear model for which the posterior computations are obtained analytically, but in other problems like in classification, sampling the parameters separately for each submodel may be very expensive, and one may be forced to use faster approximations (such as the Laplace method or expectation propagation).
On the other hand, it must be kept in mind that the possible computational savings from avoiding the construction of the full model may come at the risk of overfitting and reduced predictive accuracy, as our results show.

\section{Conclusions}
\label{sec:conclusions}

In this paper we have shortly reviewed many of the proposed methods for Bayesian predictive model selection and illustrated their use and performance in practical variable selection problems for regression and binary classification, where the goal is to select a minimal subset of input variables with a good predictive accuracy.
The experiments have been carried out using both simulated and several real world datasets.

The numerical experiments show that the overfitting in the selection may be a potential problem and hinder the model selection considerably.
This is the case especially when the dataset is small (high variance in the utility estimates) and the number of models under comparison large (large number of variables).
Especially vulnerable methods for this type of overfitting are CV, WAIC, DIC and other methods that rely on data reuse and have therefore relatively high variance in the utility estimates.
From the predictive point of view, better results are generally obtained by accounting for the model uncertainty and forming the full encompassing (reference) model with all the variables and best possible prior information on the sparsity level.
Our results showed that Bayesian model averaging (BMA) over the candidate models yields often the best results on expectation, and one should not expect to do better by selection.
This agrees with what is known about the good performance of the BMA \citep{hoeting1999,raftery2003}.

If the full model is too complex or the cost for observing all the variables is too high, the model can be simplified most robustly by the projection method which is considerably less vulnerable to the overfitting in the selection.
The advantage of the projection approach comes from taking into account the uncertainty in forming the full encompassing model and then finding a simpler model which gives similar answers as the full model.
Overall, the projection framework outperforms also the selection of the most probable variables or variable combination (Median and MAP models) being able to best retain the predictive ability of the full model while effectively reducing the model complexity.
The results also demonstrated that the projection does not only outperform the other methods on average but the variability over the different data realizations is also considerably smaller compared to the other methods.
In addition, the numerical experiments showed that constructing the submodels by the projection from the full model may improve the predictive accuracy even when some other strategy, such as marginal probabilities, are used to rank the variables.

Despite its advantages, the projection method has the inherent challenge of forming the reference model in the first place.
There is no automated way of coming up with a good reference model which emphasizes the model critisism.
However, as already stressed, incorporating the best possible prior information into the full encompassing model is formally the correct Bayesian way of dealing with the model uncertainty and often seems to also provide the best predictions in practice.
In this study we used the model averaging over the variable combinations as the reference model, but similar results are obtained also with the hierarchical shrinkage prior (see Appendix~\ref{app:proj_with_hs}).

Another issue is that, even though the projection method seems the most robust way of searching for good submodels, the estimated discrepancy between the reference model and a submodel is in general an unreliable indicator of the predictive performance of the submodel.
In variable selection, this property makes it problematic to decide how many variables should be selected to obtain predictive performance close to the reference model, even though the minimization of the discrepancy from the reference model typically finds a good search path through the model space.

However, the results show that this problem can be solved by using cross-validation outside the searching process, as this allows studying the tradeoff between the number of included variables and the predictive performance, which we believe is highly informative.
Moreover, we demonstrated that selecting the number of variables this way does not produce considerable overfitting or selection induced bias in the utility estimate for the finally selected model, because the selection is conditioned on a greatly reduced number of models (see Sec.~\ref{sec:modelsize_selection}).
While this still leaves the user the responsibility of deciding the final model size, we emphasize that this decision depends on the application and the costs of the inputs.
Without any costs for the variables, we would simply recommend using them all and carrying out the full Bayesian inference on the model.

\section*{Acknowledgements}

We thank Arno Solin and Tomi Peltola for their valuable comments to improve the manuscript. We also acknowledge the computational resources provided by Aalto Science-IT project.

\bibliographystyle{apalike}      %
\bibliography{references}   %

\begin{appendices}

\section{Projection for the linear Gaussian model}
\label{app:proj_lgm}

Consider the single output linear Gaussian regression model with several input variables \eqref{eq:model_reg}.
For this model, the projected parameters \eqref{eq:projection} can be calculated analytically.
Assume now that the reference model $M_*$ is the full model with all the inputs with any prior on the weights $w$ and noise variance $\sigma^2$.
Given a sample $(\vc w,\sigma^2)$ from the posterior of the full model, the projected parameters are given by \citep[see the derivation in][]{piironen2015b}
\begin{align}
	\label{eq:proj_weight}
	\vc w_\perp &= ({\vc X_\perp}^\tp \vc X_\perp)^{-1} {\vc X_\perp}^\tp  \vc X \vc w,  \\
	\label{eq:proj_noise}
	\sigma_\perp^2 &= \sigma^2 + \frac{1}{n} (\vc X \vc w- \vc X_\perp \vc w_\perp)^\tp (\vc X \vc w- \vc X_\perp \vc w_\perp),
\end{align}
where $\vc X = (\vc x_1^\tp,\dots,\vc x_n^\tp)$ denotes the $n\times p$ predictor matrix of the full model, and $\vc X_\perp$ the contains those columns of $\vc X$ that correspond to the submodel $M$ we are projecting onto.
The associated KL-divergence (for this particular sample) is given by
\begin{align}
	d(\vc w,\sigma^2) = \frac{1}{2}\log \frac{\sigma^2_\perp}{\sigma^2}.
\label{eq:proj_kl}
\end{align}

The projection equations~\eqref{eq:proj_weight} and \eqref{eq:proj_noise} have a nice interpretation.
The projected weights~\eqref{eq:proj_weight} are determined by the maximum likelihood solution with the observations $\vc y$ replaced by the fit of the full (reference) model $\vc f = \vc X \vc w$.
The projected noise variance~\eqref{eq:proj_noise} is the noise level of the full model plus the mismatch between the reference and the projected model.

As discussed in Section~\ref{sec:projection}, we draw a sample $\{\vc w_s,\sigma^2_s\}_{s=1}^S$ from the posterior of the reference (full) model, compute the projected parameters $\{\vc w_{s,\perp},\sigma^2_{s,\perp}\}_{s=1}^S$ and associated KL-divergences according to Equations~\eqref{eq:proj_weight}, \eqref{eq:proj_noise} and \eqref{eq:proj_kl}, and then estimate the discrepancy between the full and submodel as
\begin{align}
	\delta(M_*\|M) = \frac{1}{S} \sum_{s=1}^S d(\vc w_s,\sigma_s^2)\,.
\label{eq:proj_lgm_discrepancy}
\end{align}
For a given number of variables, we then seek for a variable combination that gives minimal discrepancy.
This procedure will produce a parsimonious model with exactly zero weights for the variables that are left out and predictive distribution similar to the full model.
The predictive distribution of the submodel \eqref{eq:projection_pred} is given by
\begin{align}
	p(\ti y \given \ti x, D, M) = \frac{1}{S} \sum_{s=1}^S \Normal{\ti y \,\given\,  {\vc w_{s,\perp}}^\tp \vc \ti x, \, \sigma_{s,\perp}^2 }.
\end{align}

\section{Projection with hierarchical shrinkage prior}
\label{app:proj_with_hs}

Here we will briefly illustrate that the projection approach (Sec.~\ref{sec:projection}) can also be used with other priors than the spike-and-slab (corresponding to the Bayesian model averaging, BMA) that we used in the main experiments in Section~\ref{sec:results}.

\subsection{Hierarchical shrinkage prior}
\label{app:hier_shrinkage}

Consider again the linear Gaussian regression model with several inputs~\eqref{eq:model_reg}.
A hierarchical shrinkage (HS) prior for the regression weights $\vc w = (w_1,\dots,w_p)$ can be obtained as
\begin{align}
\begin{split}
	w_i \given \lambda_i,\tau &\sim \Normal{0,\lambda_i^2 \tau^2}, \\
	\lambda_i &\sim \halfStudent[\nu]{0,1} \,,
\end{split}
\label{eq:hs}
\end{align}
where  $\halfStudent[\nu]{\cdot}$ denotes the half-Student-$t$ prior with $\nu$ degrees of freedom.
Intuitively, we expect the local variance parameters $\lambda_i^2$ to be large for those inputs that have large weight, and small for those with negligible weight, while the global variance term $\tau^2$ adjusts the overall sparsity level.
The shrinkage property of the prior~\eqref{eq:hs} comes from the fact that the half-$t$ prior evaluates to a positive constant at the origin and places therefore probability mass for small values of $\lambda_i$.
Moreover, if the tails of the half-$t$ densities are heavy enough, they allow some of the weights to remain unshrunk.
The horseshoe prior \citep{carvalho2009,carvalho2010} is obtained by setting $\nu=1$, that is, by introducing half-Cauchy priors for the local scale parameters $\lambda_i$.
The horseshoe prior has been shown to possess desirable theoretical properties and performance close to the gold standard BMA in practice \citep{carvalho2009,carvalho2010,datta2013,varDerPas2014}.
However, it is observed that when using Stan \citep{stan_software} for fitting the model, the NUTS sampler can produce a lot of divergent transitions even after the warm-up period \citep{piironen2015b}.
Therefore we use $\nu=3$ which is observed to behave numerically well and yield good results \citep[see][ for more details]{piironen2015b}.

\subsection{Crime dataset revisited}
\label{app:crime_with_hs}

Let us now revisit the Crime dataset from Section~\ref{sec:results_real}, which contains $1992$ observations and $p=102$ predictor variables.
We split the data randomly into two so that $n=1000$ points are used for model training and variable selection, and the remaining $\ti n=992$ points are used for testing.
We apply the regression model~\eqref{eq:model_reg} with the HS prior \eqref{eq:hs} with $\nu=3$ for the weights in the full model.
Note that we use these hierarchical priors only for the weights of the nonconstant inputs.
For the intercept term we use a weakly informative prior
\begin{align*}
	w_0 \sim \Normal{0,5^2}.
\end{align*}
For the global scale parameter $\tau$ and for the noise variance $\sigma^2$, we use the following uninformative priors
\begin{align*}
	\tau &\sim \halfCauchy{0,1}, \\
	\sigma^2 &\propto 1 \,,
\end{align*}
where $\halfCauchy{\cdot}$ denotes the half-Cauchy distribution.
The full model is fitted by drawing $S=4000$ samples from the posterior using Stan (4 chains, 2000 samples per each, first halfs discarded as a warmup).

After fitting the full model, we performed the forward variable selection as in Section~\ref{sec:results} using the projection predictive method.
In other words, by starting from the empty model, at each step we add the variable that decreases the discrepancy \eqref{eq:proj_lgm_discrepancy} to the full model the most.
The performance of the found models was then studied on the test set.
For illustration, we also cross-validated the selection within the training data.
In other words we repeated the model fitting and selection $K=10$ times each time leaving $n/K$ points out for validation, and evaluated the performance of the full model and the found submodels using these left-out data.
This was done to illustrate that the cross-validation gives a reliable estimate of the generalization performance for a given number of variables if the whole search process is cross-validated (as discussed in Sec.~\ref{sec:modelsize_selection}).

\begin{figure}[t]
\centering
	\setlength{\figureheight}{0.2\textwidth}
	\setlength{\figurewidth}{0.8\textwidth}
	\pgfplotsset{
	compat=newest,
	x tick label style={font=\footnotesize},
	y tick label style={font=\footnotesize},
	x label style={font=\footnotesize},
	y label style={font=\footnotesize},
	axis on top,
	}
	\minput[pdf]{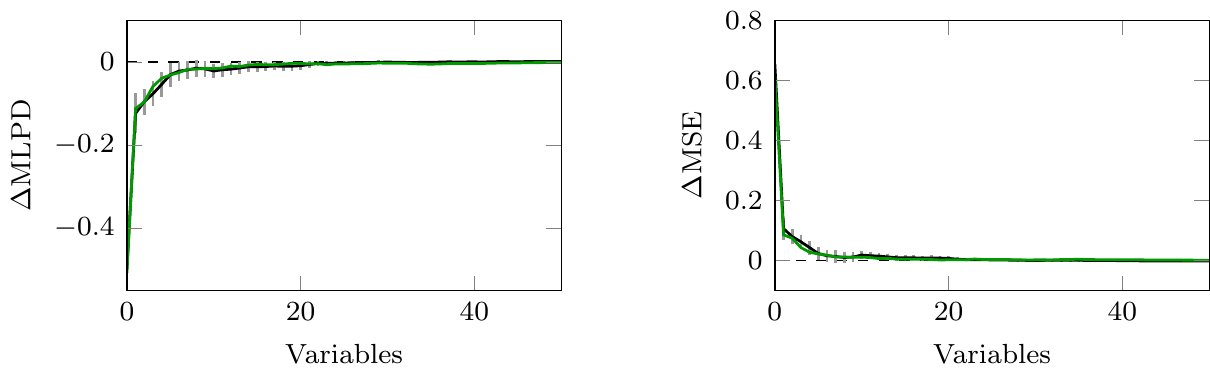}
	\caption{Crime data with HS prior for the full model: Difference in the mean log predictive density (MLPD) and mean squared error (MSE) between the projected submodel and the full model as a function of number of chosen variables up to 50 variables. Black is the average over the $K=10$ cross-validated searches within $n=1000$ data points, grey bars denote the 95\% credible interval, and green is the test performance on the remaining $\ti n=992$ test points when the search is done using all the $n=1000$ training data points.}
	\label{fig:crime_with_hs}
\end{figure}

Figure \ref{fig:crime_with_hs} shows the difference in the mean log predictive density and mean squared error between the projected submodel and the full model as a function of number of added variables up to 50 variables.
The black line is the average over the $K=10$ cross-validation folds and the green line shows the result when the fitting and searching is performed using all the training data and the performance is evaluated on the test data.
As expected, there is a good correspondence between the cross-validated and test performance.
For this dataset, most of the predictive ability of the full model is captured with about 5 variables, and 20 variables are enough for getting predictions indistinguishable from the full model for all practical purposes.
These results are essentially the same that were obtained using the BMA as the reference model (Sec.~\ref{sec:results_real}), that is, the spike-and-slab prior for the full model.

\end{appendices}

\end{document}